\documentclass[onecolumn,a4paper,accepted=2023-03-28]{quantumarticle}
\pdfoutput=1
\usepackage[utf8]{inputenc}
\usepackage{typearea}
\usepackage{amsmath,physics,algorithm,algcompatible,amssymb}
\usepackage{qcircuit}
\usepackage{graphicx}
\usepackage{authblk}
\usepackage{caption}
\usepackage{subcaption}
\usepackage{multirow}
\usepackage{color}
\usepackage{cite}
\numberwithin{equation}{section}

\algnewcommand\INPUT{\item[\textbf{Input:}]}%
\algnewcommand\OUTPUT{\item[\textbf{Output:}]}%
\algnewcommand\RETURN{\item[\textbf{return}]}%
\renewcommand{\var}[0]{{\rm Var}}

\newcommand{\Deltahybrid}[0]{\Delta_{\rm hybrid}}
\newcommand{\Deltashadow}[0]{\Delta_{\rm shadow}}
\newcommand{\Deltanaive}[0]{\Delta_{\rm naive}}
\newcommand{\gm}[0]{G_{k,i,\ell,i^{\prime},q}}
\newcommand{\gv}[0]{G_{k,i,r}}
\newcommand{\gvprime}[0]{G_{k,i,r^{\prime}}}
\newcommand{\gvu}[0]{G_{k,i,u}}
\newcommand{\gvuprime}[0]{G_{k,i,u^{\prime}}}
\newcommand{\gpp}[0]{g(\tilde{P}_r, \tilde{P}_{r^{\prime}})}
\newcommand{\gppu}[0]{g(\tilde{P}_u, \tilde{P}_{u^{\prime}})}
\newcommand{\sumru}[0]{\sum_{r = 1}^{\nba}\sum_{r^{\prime}\neq r}\sum_{u = 1}^{\nba}\sum_{u^{\prime}\neq u}}
\newcommand{\di}[0]{D_I\left(\ket{v(\vecthetaidealT)}, \ket{v(\vecthetaT)}\right)}
\newcommand{\vki}[0]{\mathcal{V}_k^{i}}
\newcommand{\rhom}[0]{\rho^{\phi_{k,i,\ell,i^{\prime}}}_{k,i;\ell,i^{\prime}}}
\newcommand{\rhov}[0]{\rho^{\phi_{k,i}}_{k,i}}
\newcommand{\numnaive}[0]{N_{\rm naive}}
\newcommand{\numparams}[0]{N_P}
\newcommand{\numderiv}[0]{N_g}
\newcommand{\nbb}[0]{N_{\rm BB}}
\newcommand{\nba}[0]{N_{\rm BA}}
\newcommand{\deldottheta}[0]{\delta{\vec{ \dot{\theta}}}(t)}

\newcommand{\vecthetaidealT}[0]{\vec{\theta}_{\rm ideal}(T)}
\newcommand{\vecthetaT}[0]{\vec{\theta}(T)}
\newcommand{\ns}[0]{N_{\rm shot}}
\newcommand{\qfunc}[1]{q(#1)}
\newcommand{\vket}[0]{\ket{v(t)}}
\newcommand{\expect}[1]{\mathbf{E}[#1]}
\newcommand{\vecthetat}[0]{\vec{\theta}(t)}
\newcommand{\vthetaket}[0]{\ket{v(\vecthetat)}}
\newcommand{\vthetabra}[0]{\bra{v(\vecthetat)}}
\newcommand{\mcnorm}[1]{\left|\left|#1 \right|\right|}
\newcommand{\locality}[1]{{\rm locality}(#1)}
\newcommand{\gkir}[0]{\mathcal{G}_{k}^{i,r}}

\newcommand{\deriv}[2]{
\frac{\partial #1}{\partial #2}
}
\newcommand{\fig}[4]{
\begin{figure}[ht]
	\includegraphics[width=#2]{#1}
	\caption{#3}
	\label{#4}
\end{figure}
}

\newcommand{\figfour}[8]{
\begin{figure}[ht]
    \begin{tabular}{cc}
    (a) & (b) \\
	\begin{subfigure}[b]{#5}
	\centering
	\includegraphics[width=\textwidth]{#1}
	\end{subfigure} &
	\begin{subfigure}[b]{#6}
	\centering
	\includegraphics[width=\textwidth]{#2}
	\end{subfigure} \\
	(c) & (d) \\
	\begin{subfigure}[b]{#5}
	\centering
	\includegraphics[width=\textwidth]{#3}
	\end{subfigure} &
	\begin{subfigure}[b]{#6}
	\centering
	\includegraphics[width=\textwidth]{#4}
	\end{subfigure}
	\end{tabular}
		\caption{#7}
	\label{#8}
\end{figure}
}

\typearea{13}
\title{Measurement optimization of variational quantum simulation by classical shadow and derandomization}

\author[1,4]{Kouhei Nakaji}

\author[2]{Suguru Endo}

\author[1]{Yuichiro Matsuzaki}

\author[3]{Hideaki Hakoshima}

\affil[1]{Device Technology Research Institute,  National Institute of Advanced  Industrial  Science and Technology  (AIST),1-1-1  Umezono,  Tsukuba,  Ibaraki  305-8568,  Japan.}
\affil[2]{NTT Computer and Data Science laboratories, NTT corporation, Musashino, Tokyo 180-8585, Japan}
\affil[3]{Center for Quantum Information and Quantum Biology, Osaka University, 1-2 Machikaneyama, Toyonaka, Osaka 560-0043, Japan.}
\affil[4]{%
Current address:
    Department of Computer Science, University of Toronto, Toronto, Ontario, Canada}

\begin{document}

\maketitle

\begin{abstract}
Simulating large quantum systems is the ultimate goal of quantum computing. Variational quantum simulation (VQS) gives us a tool to achieve the goal in near-term devices by distributing the computation load to both classical and quantum computers. However, as the size of the quantum system becomes large, the execution of VQS becomes more and more challenging. 
One of the most severe challenges is the drastic increase in the number of measurements; for example, the number of measurements tends to increase by the fourth power of the number of qubits in a quantum simulation with a chemical Hamiltonian. 
This work aims to optimize measurements in VQS by recently proposed measurement optimization techniques such as classical shadow and derandomization. Even though previous literature shows that the measurement optimization techniques successfully reduce measurements in the variational quantum optimization (VQO), how to apply them to VQS was unclear due to the gap between VQO and VQS in measuring observables. In this paper, we bridge the gap by changing the way of measuring observables in VQS and propose an algorithm to optimize measurements in VQS. 
Our numerical experiment shows the validity of using our algorithm with quantum chemical systems. 
Importantly, we theoretically reveal the advantage of using shadow-based strategies, e.g., classical shadow and derandomization, not only in VQS but also in VQO by calculating the variance of an observable. To the best of our knowledge, the advantage of using shadow-based strategies in VQO was given
only
numerically in previous research; therefore, this paper also gives significant implications for measurement optimization in general variational quantum algorithms. 
\end{abstract}


\section{Introduction}
\label{section:introduction}
With the advent of noisy intermediate scale quantum (NISQ) computers \cite{preskill2018quantum}, the next milestone is to realize their practical applications. A class of quantum algorithms called hybrid quantum-classical algorithms (see \cite{endo2021hybrid,cerezo2021variational}) came to attention because a load of computation can be distributed between quantum computers and classical computers; accordingly, the depth of the quantum circuit may be reduced. Among hybrid algorithms, variational quantum algorithms (VQAs) \cite{farhi2014quantum,peruzzo2014variational,Kandala2017,moll2018quantum,mcclean2016theory,Li2017} have been extensively studied for diverse applications such as quantum computational chemistry \cite{peruzzo2014variational,Kandala2017}, quantum machine learning \cite{mitarai2018quantum}, quantum metrology \cite{kaubruegger2019variational,koczor2020variational}, etc. In VQAs, parametrized quantum states are generated from shallow quantum circuits, and the parameters are iteratively updated with the optimization using classical computers. 
VQAs can be broadly classified into two types: variational quantum optimization (VQO) \cite{farhi2014quantum,peruzzo2014variational,Kandala2017,moll2018quantum,mcclean2016theory} and variational quantum simulation (VQS) \cite{Li2017}. VQO computes and optimizes a problem-specific cost function; for example, in the famous variational quantum eigensolver (VQE) \cite{peruzzo2014variational,Kandala2017}, the energy of parametrized quantum states is evaluated by measuring Pauli operators, which constitutes the problem Hamiltonian in quantum computers, and the minimization of the energy results in the ground state and its energy. Meanwhile, VQS simulates the dynamics of the quantum system, e.g., quantum real and imaginary time evolution \cite{Li2017,mcardle2019variational} and open quantum system dynamics \cite{yuan2019theory} by minimizing the distance between the variationally simulated evolution and the exact evolution. The update rule of the parameters in VQS is derived from variational principles for time evolution, such as McLachlan’s variational principle \cite{McLachlan1964}. 
In both VQO and VQS, updates of the parameters are performed according to the output from quantum computers, more specifically, expectation values of observables. 
Since the number of circuit runs is still restricted in the current NISQ era, it is crucial to reduce the effect of shot noise and save the measurement counts for estimating each observable.

However, with the larger scale of the problem, reducing the effect of shot noise becomes more and more challenging. That is because the number of measurements to realize a given shot noise linearly increases with the number of the terms in the observable (here, we call `term' by an observable evaluatable with a quantum circuit). 
For example, given the number of qubits as $n$, the number of terms scales as $O(n^4)$ in the chemical Hamiltonian, and therefore, the number of measurements also scales as $O(n^4)$. 
To mitigate the challenge, several measurement optimization methods are proposed in the context of VQO \cite{mcclean2016theory,Verteletskyi2020,Huang2020,Hadfield2022,Huang2021,Hillmich2021,Wu2021,gokhale2019minimizing,izmaylov2019unitary,crawford2021efficient,huggins2021efficient,hamamura2020efficient,bravyi2017tapering,zhao2020measurement,yen2020measuring,jena2019pauli,yen2022deterministic,choi2022improving,choi2023fluid}. 
In general, there are two types of measurement optimization methods: the grouping strategy \cite{mcclean2016theory,Verteletskyi2020,Hadfield2022,gokhale2019minimizing,izmaylov2019unitary,crawford2021efficient,huggins2021efficient,hamamura2020efficient,bravyi2017tapering,zhao2020measurement,yen2020measuring,jena2019pauli,yen2022deterministic,choi2022improving,choi2023fluid} and the shadow-based strategy \cite{Huang2020,Hadfield2022,Huang2021,Hillmich2021,Wu2021}. Previous results numerically show that the applications of grouping strategies \cite{hamamura2020efficient,arrasmith2020operator} and shadow-based strategies \cite{Huang2020,Huang2021,boyd2022training} to VQE of specific problems achieve faster or better convergence in terms of the number of measurements though the performance of the optimization depends on which we choose an optimizer. Let us briefly review those two types of strategies in the following. 

On the one hand, the grouping strategy is a method to efficiently estimate the cost function of VQO, such as a molecular Hamiltonian, by grouping commuting terms in the Hamiltonian \cite{mcclean2016theory,Verteletskyi2020,Hadfield2022,gokhale2019minimizing,izmaylov2019unitary,crawford2021efficient,huggins2021efficient,hamamura2020efficient,bravyi2017tapering,zhao2020measurement,yen2020measuring,jena2019pauli,Wu2021,yen2022deterministic,choi2022improving,choi2023fluid}.
We can reduce the number of measurements by simultaneously measuring all the terms in each group. 
On the other hand, 
the shadow-based strategies construct a classical representation of a quantum state with a small number of measurements \cite{Huang2020,Hadfield2022,Huang2021,Hillmich2021,Wu2021}. The classical shadow, which is the original shadow-based strategy,  predicts $M$ observables with $O(\log{M})$ measurements up to the additive error; the information-theoretic arguments show that the order of $O(\log{M})$ measurements is asymptotically optimal \cite{Huang2020}. However, the classical shadow is inefficient for estimating particular observables since it typically contains unnecessary measurement outcomes not used for estimation. Previous studies propose variations of classical shadow to predict specific observables, such as many-body Hamiltonians, including the sum of Pauli terms \cite{Hadfield2022,Huang2021,Hillmich2021,Wu2021}. One of its variations, derandomization, can efficiently predict such observables \cite{Huang2021}. Derandomization is a deterministic protocol to minimize the confidence bound of Pauli observables starting from completely randomized Pauli measurements, and it typically achieves better performance than the grouping strategy and original classical shadow. These measurement protocols are experiment-friendly even for NISQ devices, and a photonic system realized an experiment of these protocols \cite{zhang2021experimental}. 

Even though the above measurement optimization strategies mainly target VQO, measurement optimization is arguably more important in VQS than in VQO. In VQO, the evolution of parameters in each step does not need to be correct as far as the optimal value of the observable is found. Meanwhile, in the case of VQS, we expect that the evolution of parameters in each step correctly describes the time evolution, and therefore, we need to efficiently reduce the shot noise by optimizing the measurements. 
Still, there is no previous research that applies the above measurement optimization methods to VQS. As far as our knowledge, that is because the above
measurement optimization methods are mainly constructed for the problem where their observables are written by the sum of Pauli terms. In VQS, we need to evaluate the real/imaginary part of the expectation of operators \cite{Li2017,mcardle2019variational,yuan2019theory}, and cannot be straightforwardly written in the form of estimating the sum of Pauli observables. Approaches to optimize measurements in VQS are highly demanded.

The aim of this paper is to show the way of applying measurement optimization techniques to VQS. We particularly focus on shadow-based strategies. 
Our contribution also exists in clarifying the advantage of utilizing the shadow-based methods both in VQS and VQO from the theoretical aspect. As far as our knowledge, the advantage of VQO was shown only numerically in previous research. In this paper, we show the advantage by evaluating the variance of an observable. Let us specify the contribution of our work is summarized as follows:
\begin{itemize}
    \item 
We successfully simplify the quantum circuit for VQS by a circuit synthesis, so that measurement optimization techniques (shadow-based strategies and grouping strategies) are applicable,
\item We theoretically show the advantage of using the shadow-based strategy in the VQS by evaluating the variance of the observable; the analysis is also applicable for assessing the merit of using the shadow-based strategy in VQO,
\item  We show the validity of our algorithm and the theoretical discussion by numerical demonstration. We show that the derandomization \cite{Huang2021}, one of the shadow-based strategies, greatly reduces the number of measurements compared to the original approach.
\end{itemize}
    We note that even though we focus on the shadow-based methods in this paper, we can also utilize the grouping strategies for measurement optimization in VQS due to our circuit synthesis. Particularly, the recent result \cite{yen2022deterministic} shows
    that they can outperform derandomization in VQE by optimizing parameters in the grouping so that the variance of the observable is minimized. Also, in \cite{choi2022improving,choi2023fluid}, authors provide methods that may outperform the method in
    \cite{yen2022deterministic}. 
Thus, we expect better performance in VQS by using those grouping methods than shadow-based methods. It is still worth mentioning that those grouping methods need a reference quantum state for calculating the variance; the reference state is chosen to be a quantum state close to the ground state of the Hamiltonian in their work since they mainly target VQE. However, in most cases of VQS, our goal is not to find the ground state of the Hamiltonian, and we need to investigate the choice of a reference state. We leave combining our methods with those grouping strategies for future work.

The rest of the paper is organized as follows. In Section~\ref{section:VQS}, we give a brief review of VQS. We also state the challenge of VQS in the same section. We review the shadow-based strategy in Section~\ref{section:shadow-based-strategy}; particularly, we discuss the classical shadow \cite{Huang2020} and the derandomization \cite{Huang2021} in detail. Section~\ref{section:our-algorithm} is dedicated to proposing our algorithm that applies the shadow-based strategy to VQS. The advantage of utilizing the shadow-based strategy is discussed in the same section. In Section~\ref{section:numerical-demonstration}, we numerically demonstrate the performance of our algorithm. Finally, we conclude our paper with some discussions in Section~\ref{section:conclusion}.

\section{Variational quantum simulation without measurements optimization}
\label{section:VQS}
In this section, we review the variational quantum simulation (VQS) proposed and studied in the previous literature  \cite{Li2017,mcardle2019variational,yuan2019theory} and state a challenge of the algorithm. In Section~\ref{section:general-process}, we give a general description of the VQS based on \cite{Endo2020}. In Section~\ref{section:ite-rte}, we show the variational real and imaginary time evolution as examples. Section~\ref{section:problem} is dedicated to stating a problem of the previous approach in the VQS.

\subsection{General processes}
\label{section:general-process}

We define the quantum time evolution 
for a state $\vket$ in this paper as the following differential equation, 
\begin{equation}
\label{eq:general-process}
	B(t)\frac{d}{dt}\vket = A(t)\vket.
\end{equation}
where $A(t)$ and $B(t)$ denote matrices.
In the VQS, we parameterize $\vket$ as $\vthetaket$ by using a set of parameters. From Mclachlan's principle \cite{McLachlan1964}, 
\begin{equation}
\label{eq:mclachlan}
	\delta \mcnorm{
		B(t) \sum_{k=1}^{\numparams} \deriv{\vthetaket}{\theta_k}\dot{\theta_k}
		- A(t) \vthetaket
	}
	= 0,
\end{equation}
where $\mcnorm{\ket{\psi}} \equiv \braket{\psi}$, $\numparams$ is the number of parameters, and 
\begin{equation}	
\label{eq:delf}
	\delta f \equiv \sum_{j=1}^{\numparams}\deriv{f}{\dot{\theta}_j} \delta\dot{\theta_j}.
\end{equation}
We note that there are two other conventional variational principles than the Mclachlan's principle \cite{McLachlan1964}: Dirac and Frenkel's variational principles \cite{dirac_1930,frenkel1935wave} and the time-dependent variational principle \cite{kramer1981geometry,broeckhove1988equivalence}. A detailed comparison of different variational principals is made in \cite{yuan2019theory}.
The Dirac and Frenkel variational principle is not suitable for VQS since the equation of the parameters may involve complex solutions, which contradicts the requirement that parameters are real. Even though the time-dependent variational principle gives a real solution, the time-dependent variational principle is shown to be more unstable and not applicable to the evolution of density matrices and imaginary time evolution. In contrast,
McLachlan’s principle generally produces stable solutions, and it is also applicable to all VQS problems.

Equation \eqref{eq:mclachlan} is equivalent to
\begin{align}
\label{eq:mclachlan-explicit}
	\sum_{\ell=1}^{\numparams} \left[\deriv{\vthetabra}{\theta_k}B^{\dagger}(t)B(t)
	\deriv{\vthetaket}{\theta_\ell}
	+ 
	\deriv{\vthetabra}{\theta_\ell}B^{\dagger}(t)B(t)
	\deriv{\vthetaket}{\theta_k} 
	\right] \dot{\theta}_{\ell} \nonumber\\
	- \left[
	\deriv{\vthetabra}{\theta_{k}}B^{\dagger}(t)A(t)\vthetaket + \vthetabra A^{\dagger}(t)B(t)
	\deriv{\vthetaket}{\theta_k}
	\right]
	= 0.
\end{align}
for all $k$.
We can further simplify \eqref{eq:mclachlan-explicit} as
\begin{equation}
\label{eq:mclachlan-matrix-form}
	\sum_{\ell=1}^{\numparams} M_{k\ell}\dot{\theta}_{\ell} = V_k,
\end{equation}
where 
\begin{align}	
	M_{k\ell} &= \real\left[\deriv{\vthetabra}{\theta_k}B^{\dagger}(t)B(t)
	\deriv{\vthetaket}{\theta_\ell}\right], \label{eq:matrix-mkl}\\ 
	V_k &=\real\left[ 	\deriv{\vthetabra}{\theta_{k}}B^{\dagger}(t)A(t)\vthetaket, \right] \label{eq:vector-vk}
\end{align}
with $\real[w]$ as the real part of the complex number $w$. 

Once we compute $M_{k\ell}$ and $V_k$, we can update the parameter by the Euler method with a small parameter $\delta t$ as 
\begin{equation}	
\vec{\theta}(t + \delta t) =  \vec{\theta}(t) + M^{-1}
V \cdot \delta t, 
\end{equation}
where $M$ ($V$) is the matrix (vector) representation of $M_{k\ell}$ ($V_k$). Next, we show how to compute $M_{k\ell}$ and $V_k$ in the following.

\subsubsection*{Estimation of $M_{k\ell}$ and $V_k$ using quantum circuits}
Let $R$ be the unitary operator corresponding to the parameterized quantum circuit (PQC); namely 
\begin{equation}
\vthetaket = R |v_{\rm ref}\rangle, 	
\end{equation}
with $|v_{\rm ref}\rangle$ as a reference state. We often call $R$ the {\it ansatz}. 
We expand $R$ as 
\begin{equation}
\label{eq:ansatz}
	R \equiv R_{N_P}(\theta_{\numparams})R_{\numparams -1}(\theta_{\numparams-1})\cdots R_1(\theta_1),
\end{equation}
where $R_k$ is the $k$-th unitary operator applied to the reference state and $\theta_k$ is the parameter of $R_k$.
Let us describe
\begin{equation}
	\deriv{R_k(\theta_k)}{\theta_k} = \sum_{i=1}^{\numderiv} g_{k,i}R_k(\theta_k) P_{k,i},
\end{equation}
where $N_g$ is the number of Pauli observables when expanding the derivative, $g_{k,i}$ is a coefficient,  and $P_{k,i}$ is a tensor product of Pauli 
operators and identity: $P_{k,i}\in \{X, Y, Z, I\}^{\otimes n}$.
Then 
\begin{equation}
\label{eq:vderiv-expansion}
	\deriv{\vthetaket}{\theta_k} = \sum_{i=1}^{\numderiv} g_{k,i} R_{k, i}|v_{\rm ref}\rangle,
\end{equation}

where 
\begin{equation}
	R_{k,i} = R_{\numparams}(\theta_{\numparams})R_{{\numparams}-1}(\theta_{\numparams-1})\cdots 
		R_{k}(\theta_{k}) P_{k,i} R_{k-1}(\theta_{k-1})\cdots
	R_1(\theta_1). 
\end{equation}
In typical settings, $\numderiv=O(1)$ is satisfied.
For example, if we embed each parameter as 
$\exp(-i\theta_k P/2 )$, where 
$P$ denotes a tensor products of Pauli operators, $\numderiv = 1$ and $g_{k,1} = -i/2$.
Also, let us describe
\begin{align}
	B^{\dagger}(t)B(t) &= \sum_{q=1}^{\nbb} \beta_q P_q, \label{eq:bb}\\
	B^{\dagger}(t)A(t) &= \sum_{r=1}^{\nba} \alpha_r P_r, \label{eq:ba}
\end{align}
where $P_q$ and $P_r$ are tensor products of Pauli operators and identity, and where $\nbb$ and $\nba$ are the numbers of terms when decomposing $B^{\dagger}(t)B(t)$ and $B^{\dagger}(t)A(t)$ as the sums of Pauli operators.

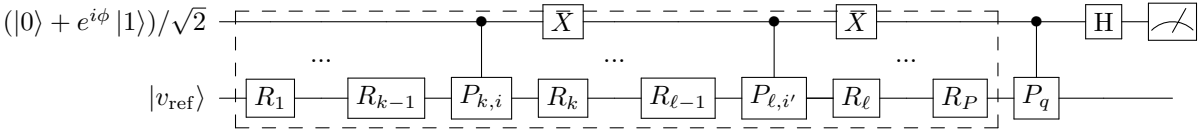
\begin{figure*}[h!]
	\centering
\begin{align*}
\Qcircuit @C=1em @R=.7em {
&&&&&&&&&& \\
\lstick{(\ket{0}+e^{i\phi}\ket{1})/\sqrt{2}}&\qw&\qw&\qw&\ctrl{2}&\gate{X}&\qw &\qw & \ctrl{2}&\gate{X} & \qw &\qw &\ctrl{2} &\gate{\rm H}& \meter\\
&&...&&&&...&&&&...\\
\lstick{\ket{v_{\rm ref}}}&\gate{R_1}&\qw&\gate{R_{k-1}}&\gate{P_{k,i}}&
\gate{R_{k}}& \qw & \gate{R_{\ell-1}} &\gate{P_{\ell,i^{\prime}}}& \gate{R_{\ell}}\qw & \qw &\gate{R_P}&\gate{P_q}&\qw&\qw 
\gategroup{2}{2}{4}{12}{.8em}{--} 
}
\end{align*}
\caption{A quantum circuit to compute $\real\left[
	e^{i\phi} \bra{v_{\rm ref}}R_{k,i}^{\dagger}P_q R_{\ell,i^{\prime}} \ket{v_{\rm ref}}\right]$ for $k \leq \ell$. We use the circuit enclosed by the dotted rectangle in Section~\ref{section:our-algorithm}. We note that the output quantum state after applying the gates in the dotted box is $|\psi_{k,i;\ell,i^{\prime}}^{\phi}\rangle = 1/\sqrt{2}\left(R_{\ell, i^{\prime}}|0\rangle |v_{\rm ref}\rangle + e^{i\phi} R_{k,i}|1\rangle|v_{\rm ref}\rangle\right)$.}
	\label{fig:circuit-mkl}
\end{figure*}

\begin{figure*}[h!]
	\centering
\begin{align*}
\Qcircuit @C=1em @R=.7em {
&&&&&&&&&& \\
\lstick{(\ket{0}+e^{i \phi}\ket{1})/\sqrt{2}}&\gate{X}&\qw&\qw&\ctrl{2}&\gate{X}&\qw&\qw&\ctrl{2}&\gate{\rm H}& \meter\\
&&...&&&&...&\\
\lstick{\ket{v_{\rm ref}}}&\gate{R_1}&\qw&\gate{R_{k-1}}&\gate{P_{k,i}}&\gate{R_{k}}&\qw&\gate{R_{P}}&\gate{P_r}&\qw&\qw
\gategroup{1}{2}{4}{8}{.8em}{--} 
}
\end{align*}

\caption{A quantum circuit to compute $\real\left[
	e^{i\phi}
	\bra{v_{\rm ref}}R_{k,i}^{\dagger}P_r R \ket{v_{\rm ref}}
	\right]$. As in Figure~\ref{fig:circuit-mkl}, we use the circuit enclosed by the dotted rectangle in Section~\ref{section:our-algorithm}.
	We note that the output quantum state after applying the gates in the dotted box is $|\psi^{\phi}_{k,i}\rangle = 1/\sqrt{2}(|0\rangle R_{k,i}|v_{\rm ref}\rangle + |1\rangle R|v_{\rm ref}\rangle)$.
	}
\label{fig:circuit-vk}
\end{figure*}
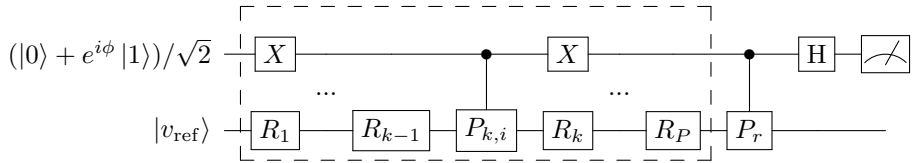

By substituting \eqref{eq:vderiv-expansion} and \eqref{eq:bb} into \eqref{eq:matrix-mkl}, we obtain 
\begin{equation}
\label{eq:mkl-real}
	M_{k\ell} = \sum_{i, i^{\prime}}^{N_g}\sum_{q=1}^{\nbb} \real\left[
	g_{k,i}^{\ast} g_{\ell,i^{\prime}}\beta_q \bra{v_{\rm ref}}R_{k,i}^{\dagger}P_q R_{\ell,i^{\prime}} \ket{v_{\rm ref}}
	\right].
\end{equation}
Similarly, by substituting \eqref{eq:vderiv-expansion} and \eqref{eq:ba} into \eqref{eq:vector-vk}, we obtain 
\begin{equation}
\label{eq:vk-real}
	V_k = \sum_{i=1}^{\numderiv}\sum_{r=1}^{\nba} \real\left[
	g_{k,i}^{\ast}\alpha_r
	\bra{v_{\rm ref}}R_{k,i}^{\dagger}P_r R \ket{v_{\rm ref}}
	\right].
\end{equation}
For computing \eqref{eq:mkl-real}, we need to evaluate each
$\real\left[
	e^{i\phi} \bra{v_{\rm ref}}R_{k,i}^{\dagger}P_q R_{\ell,i^{\prime}} \ket{v_{\rm ref}}\right]$ with $e^{i\phi}$ as a phase, which is possible by using the Hadamard test. In Fig.~\ref{fig:circuit-mkl}, we show the quantum circuit to compute the value for $k \leq \ell$. 
		Note that since $M_{k\ell} = M_{\ell k}$, we do not need to consider the case when $k > \ell$. 
To compute \eqref{eq:vk-real}, we need to evaluate $ \real\left[
	e^{i\phi}
	\bra{v_{\rm ref}}R_{k,i}^{\dagger}P_r R \ket{v_{\rm ref}}
	\right]$, which is possible by using the quantum circuit in Fig.~\ref{fig:circuit-vk}. 

\subsection{Variational real and imaginary time evolution}
\label{section:ite-rte}
Here we show $M_{k\ell}$ and $V_k$ in real-time evolution (RTE) and imaginary time evolution (ITE) as examples of the VQS. The real-time evolution is an algorithm to simulate the Schr\"{o}dinger equation under a time-evolution Hamiltonian $H$:
\begin{equation}
\label{eq:schrodinger}
	\frac{\partial |v(t)\rangle}{\partial t} = -i H |v(t)\rangle.
\end{equation}
Thus, $A(t) = -iH$ and $B(t) = 1$ and we obtain 
\begin{align}	
\label{eq:mkl-rte}
	M_{k\ell} &= \sum_{i, i^{\prime}}^{\numderiv}  \real\left[ 
 g^{\ast}_{k,i} g_{\ell,i^{\prime}} \langle v_{\rm ref}|R_{k,i}^{\dagger} R_{\ell, i^{\prime}}|v_{\rm ref}\rangle
	\right], \\
\label{eq:vk-rte}
	V_k &= \sum_{i=1}^{\numderiv}\sum_{r=1}^{\nba} \real \left[
			-i g_{k,i}^{\ast} \alpha_r \bra{ v_{\rm ref}} R_{k,i}^{\dagger} P_r R \ket{v_{\rm ref}}
	\right],
\end{align}
where we expand the Hamiltonian as $H = \sum_{j=1}^{\nba} \alpha_r P_r$ with $\alpha_r$ as a real coefficient. Note that we see $\nbb = 1$  from \eqref{eq:mkl-rte}.

The imaginary time evolution is the operation obtained by replacing $it$ to $\tau$ and $H$
to $H - \langle H \rangle$ in \eqref{eq:schrodinger}, where $\langle H \rangle = \langle v(t)| H|v(t) \rangle$. Namely, 
\begin{equation}
	\frac{\partial \ket{v(\tau)}}{\partial \tau} = - (H - \langle H \rangle) \ket{v(\tau)}.
\end{equation}
Note that the term $\langle H \rangle$ is added in order to preserve the norm of the quantum state. Importantly, we can show that, in a large $\tau$ limit, $\ket{v(\tau)}$ becomes the ground state of the Hamiltonian $H$. Thus, the imaginary time evolution is often used as a way of searching for the ground state. The functions $A(t)$ and $B(t)$ are given by $A(t) =-(H - \langle H\rangle)$ and $B(t) = 1$. The term $\langle H \rangle$ does not affect the computation of $V_k$, which can be checked by substituting the values of $A(t)$ and $B(t)$ to \eqref{eq:vector-vk} as
\begin{align}	
\label{eq:vector-vk-ite}
	V_k &= \real\left[
	 \vthetabra - (H - \langle H \rangle )
	\deriv{\vthetaket}{\theta_k}
	\right], \\
	&= -\real\left[
	 \vthetabra H \deriv{\vthetaket}{\theta_k} 
	\right]
	+ \frac{1}{2}\langle H \rangle \frac{\partial\left(\vthetabra \left.v(\vec{\theta(t)})\right\rangle\right)}{\partial t}.
\end{align}
The second term, which corresponds to $\langle H \rangle$, vanishes since the norm of $\vthetaket$ preserves under the transformation of $\vecthetat$, and therefore, we obtain $M_{k\ell}$ and $V_k$ by setting $B(t) = 1$ and $A(t) = -H$ when computing \eqref{eq:mkl-real} and \eqref{eq:vk-real}. Thus, in the variational imaginary time evolution, we obtain
\begin{align}	
\label{eq:mkl-ite}
	M_{k\ell} &= \sum_{i, i^{\prime}}^{N_g}  \real\left[ 
 g^{\ast}_{k,i} g_{\ell,i^{\prime}} \langle v_{\rm ref}|R_{k,i}^{\dagger} R_{\ell, i^{\prime}}|v_{\rm ref}\rangle
	\right], \\
	\label{eq:vk-ite}
	V_k &= -\sum_{i=1}^{N_g} \sum_{r=1}^{\nba} \real \left[
			 g_{k,i}^{\ast} \alpha_r \bra{ v_{\rm ref}} R_{k,i}^{\dagger} P_r R \ket{v_{\rm ref}}
	\right].
\end{align}
The circuit used for evaluating each term of $V_k$ in real and imaginary time evolution is Fig.~\ref{fig:circuit-vk}, which we introduced in Section~\ref{section:general-process}. On the other hand, a simpler circuit is used for evaluating $M_{k\ell} $ in real and imaginary time evolution; we show the circuit in Fig. \ref{fig:concise-circuit}. We can 
readily check that the output of the circuit returns $\real \left[e^{i\phi} \bra{v_{\rm ref}} R_{k,i}^{\dagger} R_{\ell, i^{\prime}}\ket{v_{\rm ref}}\right]$. 

\begin{figure*}[h!]
	\centering
\begin{align*}
\Qcircuit @C=1em @R=.7em {
&&&&&&&&&& \\
\lstick{(\ket{0}+e^{i\phi}\ket{1})/\sqrt{2}}&\gate{X}&\qw&\qw&\ctrl{2}&\gate{X}&\qw &\qw & \ctrl{2} &\gate{\rm H}& \meter\\
&&...&&&&...&&&&\\
\lstick{\ket{v_{\rm ref}}}&\gate{R_1}&\qw&\gate{R_{k-1}}&\gate{P_{k,i}}&
\gate{R_{k}}& \qw & \gate{R_{\ell-1}} &\gate{P_{\ell,i^{\prime}}}&\qw&\qw
}
\end{align*}
\caption{A quantum circuit to compute $\real \left[ e^{i\phi} \bra{v_{\rm ref}} R_{k,i}^{\dagger} R_{\ell, i^{\prime}}\ket{v_{\rm ref}}\right]$ for $k \leq \ell$.}
\label{fig:concise-circuit}
\end{figure*}

\subsection{Problems of the current approach}
\label{section:problem}
The approach mentioned above to performing the VQS is challenging to implement because
the number of measurements linearly increases as 
the number of terms denoted by $\nbb$ and $\nba$ increases. That is because we need to evaluate the circuit corresponding to each term inside the summations of \eqref{eq:mkl-real} and \eqref{eq:vk-real} one by one (the number of measurements also linearly increases according to the value of $\numderiv$, but it is not problematic since $\numderiv$ is typically taken to be small). For example, in RTE and ITE, the number of measurements for evaluating $V_k$ linearly scales with the number of terms in $H$, which could be impractical when $H$ has many terms. Suppose that
we use a molecular Hamiltonian as the time-evolution Hamiltonian. In that case, the number of terms in the Hamiltonian is $O(n^4)$ with $n$ as the number of qubits, and therefore, the number of measurements also increases in proportion to $O(n^4)$.

The challenge of a large number of Pauli observables in Hamiltonian is also addressed
in the variational quantum eigensolver (VQE), and many solutions have been proposed \cite{Kandala2017,bravyi2017tapering,gokhale2019minimizing,Izmaylov2019,jena2019pauli,Zhao2020,hamamura2020efficient,Yen2020,Verteletskyi2020,Torlai2020,Crawford2021,Huggins2021,Huang2020,Hadfield2022,Huang2021,Hillmich2021,Wu2021}. However, 
to our best knowledge, no previous work has explored this issue in the context of the VQS. 
Importantly, there is a crucial difference between VQS and VQE;
namely, to evaluate each term in the summation in  \eqref{eq:bb} and \eqref{eq:ba}, we need the controlled operations associated with each Pauli observable to evaluate each term in the summation of  \eqref{eq:bb} and \eqref{eq:ba} in VQS, while the VQE only requires estimating the expectation value of each Pauli observable. 
Thus, it is not straightforward to apply the methods discussed in \cite{Kandala2017,bravyi2017tapering,gokhale2019minimizing,Izmaylov2019,jena2019pauli,Zhao2020,hamamura2020efficient,Yen2020,Verteletskyi2020,Torlai2020,Crawford2021,Huggins2021,Huang2020,Hadfield2022,Huang2021,Hillmich2021,Wu2021} to VQS.
In the rest of the paper, we show a way to overcome the difficulties
and discuss how to apply the proposed methods for the VQE to the VQS. Before discussing the main point, we will review the proposed methods for the VQE in the next section. 

It should be noted that in the case of ITE, the parameter shift-rule \cite{schuld2019evaluating} is applicable to calculate $V_k$ under a suitable condition (e.g., $N_g=1$ and $g_{k,1} \neq 0$ is real), meaning that the evaluation of $V_k$ boils down to the evaluation of the expectation value of $H=\sum_{r} \alpha_r P_r$ with respect to a quantum state generated by a PQC. In that case, we can reduce the number of measurements by utilizing the techniques proposed in VQE. Still, in most of the applications of VQS, the parameter-shift rule is not applicable; therefore, we need to construct a measurement optimization procedure tailored for VQS.

\section{Measurements optimization for sum of Pauli observables}
\label{section:shadow-based-strategy}
In this section, we review strategies to reduce the number of measurements in the VQE, when the Hamiltonian has many terms such as $\mathcal{H} = \sum_{j=1}^K a_jP_j$ for $K\gg 1$. In the VQE, we need to estimate the value $\langle \mathcal{H} \rangle = \sum_{j=1}^K a_j{\rm Tr}(P_j \rho)$
where $\{a_j\}$ are real coefficients, $\{P_j\}$ are Pauli observables ($P_j\in \{X, Y, Z, I\}^{\otimes n}$), and $\rho$ is a $n$-qubit quantum state generated by a quantum circuit. When $K$ increases, the number of measurements to achieve
a given precision linearly increases as far as we naively evaluate each term.   

For improving the scaling of the number of measurements, several solutions have been proposed \cite{Kandala2017,bravyi2017tapering,gokhale2019minimizing,Izmaylov2019,jena2019pauli,Zhao2020,hamamura2020efficient,Yen2020,Verteletskyi2020,Torlai2020,Crawford2021,Huggins2021,Huang2020,Hadfield2022,Huang2021,Hillmich2021,Wu2021}. The most promising solutions are arguably shadow-based strategies \cite{Huang2020,Hadfield2022,Huang2021,Hillmich2021,Wu2021}, which we review in this section.
In Section~\ref{section:general-argument}, we first give a general discussion about shadow-based strategies \cite{Huang2020,Hadfield2022,Huang2021,Hillmich2021,Wu2021}. Next, in Section~\ref{section:cs-detail}, we show the detail of some of the shadow-based strategies that we will examine in our numerical experiment in Section~\ref{section:numerical-demonstration}.

\subsection{General discussion for shadow-based strategies}
\label{section:general-argument}
Shadow-based strategies for estimating $\mathcal{H}$ \cite{Huang2020,Hadfield2022,Huang2021,Hillmich2021,Wu2021} are mostly composed of three steps. The first step of the estimation process takes the number of measurements $\ns$,
$\{a_j\}_{j=1}^K$ and  $\{P_j\}_{j=1}^K$ as its inputs. Let us denote a measurement basis by one of the Pauli observables $\{X, Y, Z\}^{\otimes n}$, where each measurement is performed on the basis where the Pauli observable is diagonalized. For example, when $n=3$, the measurement basis can be expressed as $XZX$, meaning that the quantum state $\rho$ is measured in the computational basis after we operate the Hadamard gates to the first and the third qubits. 
The first step returns the array of measurement basis as $\{M_r\}_{r=1}^{\ns}$; how to select those basis depends on the algorithm. We define the function corresponding to the first step as ${\bf buildMeasurements}(\{a_j\}, \{P_j\}, \ns)$. The second step takes $\{M_r\}_{r=1}^{\ns}$ and $\rho$ as its inputs and returns the measurement results of $\rho$ as $\{b_{r}\}_{r=1}^{\ns}$
, 
where $b_r \in \{1, -1\}^{\otimes n}$.  We denote the function corresponding to the second step as ${\bf getMeasurementResults}(\{M_r\}_{r=1}^{\ns}, \rho)$. 
The third step takes $\{M_r\}_{r=1}^{\ns}$, $\{b_{r}\}_{r=1}^{\ns}$, $\{P_j\}_{j=1}^K$, and $\{a_j\}_{j=1}^K$ as its inputs, and returns the following $\nu$:
\begin{equation}	
	\nu = \frac{1}{\ns}  	\sum_{r=1}^{\ns} \nu_r,
\end{equation}
where
\begin{equation}	
	\nu_r = \sum_{j=1}^K a_j \frac{1}{\qfunc{P_j}} f(P_j, M_r)\mu(P_j, b_r).  
\end{equation} 
Here, $f(P, M)$ denotes a covering function, $\qfunc{P_j}$ denotes a covering probability function, and $\mu(P, b)$ denotes an estimation function.
These are defined so that $\nu$ should be an unbiased estimator of ${\rm Tr}(\mathcal{H}\rho)$. We show the detailed definitions as follows. 


\subsubsection*{Covering function $f(P, M)$}
The covering function $f(P, M)$ takes a Pauli obserable $P$ and a measurement basis $M$ as its inputs, and returns outputs as follows
\begin{align}
f(P, M) &= \prod_{i=1}^n f_{\rm local}(P[i], M[i]), \\
	f_{\rm local}(P[i], M[i]) &= \left\{\begin{array}{cl}
		1 & {\rm if}~P[i]=M[i]~{\rm or}~P[i]=I \\
		0 & {\rm otherwise}\\
	\end{array}
	\right. ,
\end{align}
where $P[i]$ and $M[i]$ is the Pauli observable corresponds to the $i$-th qubit of $P$ and $M$. For example, $f(XYI, XYX) = 1$ and $f(ZIX, ZXY) = 0$. We call $M$ {\it covers} $P$ if and only if $f(P, M) = 1$ as in \cite{Huang2021}. 

\subsubsection*{Covering probability function}
We define the covering probability function $\qfunc{P}$ depending on whether we generate the measurement basis probabilistically or not. When they are probabilistically generated according to a probability distribution $q_{\rm{m}}(M)$, $\qfunc{P}$ is the probability that the Pauli observable $P$ is covered, namely 
\begin{equation}
\label{eq:covering-probability-probabilistic}
	\qfunc{P} = \sum_{M \in \{X,Y,Z\}^{\otimes n}} q_{\rm{m}}(M) f(P, M).
\end{equation}
When measurements are chosen deterministically, we define the covering probability function by the ratio that $P$ is covered when measurements are $\{M_{r^{\prime}}\}_{r^{\prime}}^{\ns}$, namely,
\begin{equation}
\label{eq:covering-probability-deterministic}
	\qfunc{P} = \frac{1}{\ns}\sum_{r^{\prime}=1}^{\ns} f(P, M_{r^{\prime}}).
\end{equation}
Note that in this deterministic case, $\qfunc{P}$ depends on the chosen measurement basis $\{M_{r^{\prime}}\}_{r^{\prime}=1}^{\ns}$, but we omit the corresponding index from $\qfunc{P}$ for simplicity.

\subsubsection*{Estimation function}
The second function is the estimation function $\mu(P, b)$ that takes the Pauli observable and the bit array $b \in \{1, -1\}^{\otimes n}$ as its inputs and operates as follows:
\begin{align}
	\mu(P, b) &= \prod_{i=1}^n \mu_{\rm local}(P[i], b[i]), \\
	\mu_{\rm local}(P[i], b[i]) &= \left\{\begin{array}{cl}
		1 & P[i] = I~{\rm or}~b[i]=1 \\
		-1 & {\rm otherwise}
	\end{array}
	\right., 
\end{align}
where $b[i]$ is the $i$-th bit value. Let $p(\rho, M; b)$ be the probability that $n$-bit bitarray $b$ is obtained as a result of measuring $\rho$ in a basis $M$ covering $P$. Then, we can readily show
\begin{equation}
\label{eq:cover-exp}
	\sum_b p(\rho, M; b) \mu(P, b) = {\rm Tr}(P\rho).
\end{equation} 

As we show in Appendix~\ref{section:statistics}, $\nu$ becomes an unbiased estimator of ${\rm Tr}(\mathcal{H}\rho)$, i.e., $\expect{\nu} = {\rm Tr}(\mathcal{H}\rho)$. Also, the variance of $\nu$ is given by 
\begin{equation}
\label{eq:var}
		\var(\nu) =  \expect{\nu^2} - (\expect{\nu})^2  = \frac{1}{\ns}\left[
		 \sum_{j=1}^K \frac{a_j^2}{\qfunc{P_j}} + \sum_{j = 1}^K\sum_{k \neq j} a_j a_{\ell} g(P_j, P_{\ell}){\rm Tr}(P_j P_{\ell}\rho) - [{\rm Tr}(\mathcal{H}\rho)]^2
	\right], 
\end{equation} 
where if $M_r$ is sampled from $q_{\rm{m}}(M_r)$, 
\begin{equation}
\label{eq:g-probabilistic}
	g(P_j, P_{\ell}) = \sum_{M_r \in \{X, Y, Z\}^{\otimes n}} \frac{f(P_j, M_r)}{\qfunc{P_j}}  \frac{f(P_{\ell}, M_r)}{q(P_{\ell})} q_{{\rm{m}}}(M_r),  
\end{equation}
and if $\{M_r\}_{r=1}^{N_{\rm shot}}$ is deterministically chosen, 
\begin{equation}
	g(P_j, P_{\ell}) = \frac{1}{\ns}\sum_{r=1}^{\ns} \frac{f(P_j, M_r)}{\qfunc{P_j}}  \frac{f(P_{\ell}, M_r)}{\qfunc{P_{\ell}}}.   
\end{equation} 
We write the three steps of obtaining $\nu$ as {\bf estimateNu$(\rho, \{a_j\}_{j=1}^K, \{P_j\}_{j=1}^K, \ns)$} and summarize it in {\bf Algorithm~\ref{alg:estimate_h}}. It should be noted that the variance $\var(\nu)$ determines the number of measurements $N_{\rm shot}$ to achieve a given precision $\epsilon$; namely, $N_{\rm shot}$ scales as $N_{\rm shot} = O(\var(\nu)/\epsilon^2)$.
 \begin{algorithm}[ht]
  \caption{{\bf estimateNu$(\rho, \{a_j\}, \{P_j\}, \ns)$}}\label{alg:estimate_h}
  \begin{algorithmic}[1]
\INPUT $\rho, \{a_j\}, \{P_j\}, \ns$
  \STATE
  	 $\{M_r\}_{r=1}^{\ns} = $~{\bf buildMeasurements}$(\{a_j\}, \{P_j\}, \ns)$
  	\STATE Set $\{b_r\}_{r=1}^{\ns}=$~${\bf getMeasurementResults}(\{M_r\}, \rho)$
  	\FOR{$r=1...\ns$}
  	\STATE 
\begin{flalign*}	
	\nu_r = \sum_{j=1}^K a_j \frac{1}{\qfunc{P_j}} f(P_j, M_r)\mu(P_j, b_r). 
\end{flalign*}
\ENDFOR
    \RETURN $\nu = \frac{1}{\ns}\sum_{r=1}^{\ns}\nu_r$.
  \end{algorithmic}
\end{algorithm}


\subsection{Examples of shadow-based strategies }
\label{section:cs-detail}
The shadow-based strategy is originally proposed by the literature \cite{Huang2020} and various methods motivated by the technique have been proposed \cite{Huang2020,Hadfield2022,Huang2021,Hillmich2021,Wu2021}, which mainly target to reduce the value of $\var(\nu)$ in \eqref{eq:var}.
Among them, we review the classical shadow and the derandomization in the following discussion.

\begin{algorithm}[ht]
  \caption{{\bf buildMeasurements}($\{a_j\}, \{P_j\}, \ns$) in Classical shadow}\label{alg:buildMeasurements_cs}
  \begin{algorithmic}[1]
  \INPUT $\{a_j\}, \{P_j\}, \ns$
    \STATE SET $n$ as the number of qubits of the operators $\{P_j\}$.
  	\STATE SET results$~= []$. 
  	\FOR{$r=1...\ns$} 
  	    \STATE  SET $M = []$.
  		\FOR{$i=1...n$}
  	\STATE Choose one of  $\{X, Y, Z\}$ with equal probability and set it to $M[i]$.
  	  \ENDFOR
  	  \STATE SET results[$r$]$~= M$.
  	\ENDFOR
    \RETURN results
  \end{algorithmic}
\end{algorithm}

\subsubsection*{Classical shadow}
In the classical shadow, we randomly choose the measurement basis. 
Namely, each $M_r[i]$ ($i$-th Pauli matrix in $r$-th measurement basis) is chosen from $\{X, Y, Z\}$ with equal probability, i.e., $q_{\rm{m}}(M) = 1/3^n$. We show the pseudo-code for ${\bf buildMeasurements}(\ns)$ in classical shadow in Algorithm \ref{alg:buildMeasurements_cs}. 
We can readily show that the covering probability is given by
\begin{equation}
\label{eq:covering-cs}
	q(P_j) = \sum_{M \in \{X, Y, Z\}^{\otimes n} } \frac{f(P_j, M)}{3^n}  = \frac{1}{3^{{\rm locality}(P_j)}}, 
\end{equation}
where ${\rm locality}(O)$ is the number of non-identity Pauli matrix in the Pauli observable $O$. As we show in Appendix~\ref{section:variance-cs}, 
we obtain 
\begin{align}
	g(P_j, P_{\ell}) = g_{\rm cs}(P_j, P_{\ell})\equiv\prod_{i=1}^n g_{\rm local}(P_j[i], P_{\ell}[i]), 
\end{align}
where 
\begin{equation}
	g_{\rm local}(P_j[i], P_{\ell}[i]) = 
	\left\{
	\begin{array}{cl}
	1	& P_j[i] = I~{\rm or}~P_{\ell}[i] = I \\
	3 & P_j[i] = P_{\ell}[i] \neq I \\
	0 & otherwise
	\end{array}
	\right. .
\end{equation}
Therefore, 
\begin{equation}
\label{eq:var-cs}
	\var(\nu) = \frac{1}{\ns}\left[
		 \sum_{j=1}^K a_j^2 3^{{\rm locality}(P_j)}+ \sum_{j\neq \ell} a_j a_{\ell} g_{\rm cs}(P_j, P_{\ell}){\rm Tr}(P_j P_{\ell}\rho) - [{\rm Tr}(\mathcal{H}\rho)]^2
	\right].
\end{equation}
\subsubsection*{Derandomization}
As we see in \eqref{eq:var}, the value of $\var(\nu)$ tends to be small if $q(P_j)$ is large. From this point of view, the random choice of measurements in the classical shadow is not always efficient; particularly, when the locality of a Pauli term $P_j$ is large, the value of  $q(P_j)$ is exponentially suppressed, as we see in \eqref{eq:covering-cs}. Thus, strategies to choose the measurement basis set that more efficiently cover Pauli observables are proposed as consecutive works \cite{Hadfield2022,Huang2021,Hillmich2021,Wu2021}. 

One of such strategies is the derandomization \cite{Huang2021}. 
Unlike the classical shadow, a pre-defined measurement basis set is used in the derandomization (thus, it corresponds to the deterministic case). For obtaining the pre-defined measurement basis $\{M_r\}_{r=1}^{N_{\rm shot}}$, the procedure of the derandomization iteratively determines them in the following order:
\begin{equation}	
M_1[1], M_1[2],\cdots M_1[n], M_2[1], M_2[2], \cdots M_2[n], \cdots M_{N_{\rm shot}}[1], M_{N_{\rm shot}}[2], \cdots M_{N_{\rm shot}}[n].
\end{equation}
Suppose that $\{M_{r}\}_{r=1}^{r_0}$ and 
$\{M_{r_0+1}[i]\}_{i=1}^{i_0}$
are already determined and let ${\bf M}$ be the concatinate of  $\{M_{r}\}_{r=1}^{r_0}$ and $\{M_{r_0+1}[i]\}_{i=1}^{i_0}$.
 Then, $M_{r_0+1}[i_0 + 1] \in \{X, Y, Z\}$
is determined so that a score function $C({\bf M}, M_{r_0+1}[i_0 + 1])$
is maximized (here we assume $i_0 < n$ for simplicity, but we can give the same argument by replacing $M_{r_0+1}[i_0 + 1]$
with $M_{r_0+2}[1]$
even when $i_0 = n$). 

The score function $C$ is constructed so that the increase of $C$ leads to the increase of $\{q(P_j)\}_{j=1}^K$ in our notation. Let
\begin{equation}
	V(P_j, {\bf M}, M_{r_{0}+1}[i_0 + 1]) = \frac{\eta}{2} \sum_{r=1}^{r_0} f(P_j, M_r) 
	- \log\left(1 - \beta_{i_0 + 1} \prod_{k=1}^{i_0 + 1} f_{\rm local}(P_j[k], M_{r_0+1}[k]) \right)
\end{equation}
with 
\begin{equation}
	\beta_{i_0+1} = \gamma \prod_{j=i_0+2}^K \frac{1}{3^{\locality{P_j}}}, 
\end{equation} 
where $\gamma, \eta$ are hyperparameters. 
In \cite{Huang2021}, they give some options for the choice of the score function $C$; among them, the one used in their numerical experiment is 
\begin{equation}
\label{eq:dr-cost}
	C({\bf M}, M_{r_0+1}[i_0 + 1]) 
	= \sum_{j=1}^K \exp\left(-V\left(P_j, {\bf M}, M_{r_0+1}[i_0 + 1]\right)/w_j \right), 
\end{equation}
where 
\begin{equation}
	w_j = \frac{|a_j|}{\max_j |a_j|}.
\end{equation}

For example, suppose that the observable is 
\begin{equation}
\label{eq:easyHamiltonian}
	H = XXXZ + XXII +   IIXZ + YYZX + YYII + IIZX
\end{equation}
and $N_{\rm shot} = 10$. By minimizing \eqref{eq:dr-cost} step by step, $\{M_r\}_{r=1}^{10}$ includes five $XXXZ$ and $YYZX$, which leads to 
\begin{equation}
	\frac{1}{q(P_r)} = 2.
\end{equation}
In the classical shadow, 
\begin{equation}
	\frac{1}{q(P_r)} \geq 3^2.
\end{equation}
Thus, we see that the covering probability improves in the derandomization compared to the classical shadow. 

\section{Measurements optimization in the VQS}
\label{section:our-algorithm}
In this section, we propose the VQS with shadow-based strategies reviewed in the previous section. 
In Section~\ref{section:circuit-synthesis}, we synthesize the circuits in Fig.~\ref{fig:circuit-mkl} and Fig.~\ref{fig:circuit-vk} and remove unnecessary two-qubit operations.
Based on the result of the circuit synthesis, we propose an algorithm applying shadow-based strategies to the VQS in Section~\ref{section:algorithm-main}. In Section \ref{section:algorithm-advantage}, we discuss the advantage of using our proposed algorithm. Note that we generally discuss our proposed algorithm in this section and demonstrate its performance by using RTE and ITE in the numerical demonstration in Section~\ref{section:numerical-demonstration}. Section~\ref{section:remark} is dedicated to apply our implication obtained in Section~\ref{section:algorithm-advantage} to the general shadow-based strategies discussed in Section~\ref{section:shadow-based-strategy}
\subsection{Circuit synthesis}
\label{section:circuit-synthesis}
Let us define $\rho^{\phi}_{k,i;\ell,i^{\prime}}$ as the quantum state generated by operating the partial circuits enclosed by the dotted rectangle in Fig.~\ref{fig:circuit-mkl} to the reference state: $(|0\rangle + e^{i\phi}|1\rangle)/\sqrt{2} \otimes \ket{v_{\rm ref}}$. 
The specific form of $	\rho^{\phi}_{k,i;\ell,i^{\prime}}$ is given by
\begin{align}
	\rho^{\phi}_{k,i;\ell,i^{\prime}} &= |\psi_{k,i,\ell;i^{\prime}}^{\phi}\rangle \langle\psi_{k,i;\ell,i^{\prime}}^{\phi}|,  \\
|\psi_{k,i;\ell,i^{\prime}}^{\phi}\rangle &= \frac{1}{\sqrt{2}}\left(R_{\ell, i^{\prime}}|0\rangle |v_{\rm ref}\rangle + e^{i\phi} R_{k,i}|1\rangle|v_{\rm ref}\rangle\right).
\end{align}
Let ${\rm CP}_q$ also be the controlled-$P_q$ operation and $I_n$ be the $n$-qubit identity operator. 
Then, for each $q$, $k$, and $\ell$, the output of the circuit shown in Fig.~\ref{fig:circuit-mkl} has the following form, 
\begin{align}	
\real\left[
	e^{i\phi} \bra{v_{\rm ref}}R_{k,i}^{\dagger}P_q R_{\ell,i^{\prime}} \ket{v_{\rm ref}}\right] &= {\rm Tr}\left[ (X \otimes I_n) ({\rm CP}_q) \rho^{\phi}_{k,i;\ell,i^{\prime}} ({\rm CP}_q^{\dagger}) \right] \\
	&= 	{\rm Tr}\left(O_q \rho^{\phi}_{k,i;\ell,i^{\prime}} \right),   
\end{align}
with 
\begin{equation}
\label{eq:oq}
	O_q \equiv ({\rm CP}_q^{\dagger}) (X \otimes I_n) ({\rm CP}_q).
\end{equation}
Similarly, let $\rho^{\phi}_{k,i}$ be the quantum state generated by operating the partial circuits enclosed by the dotted rectangle in Fig.~\ref{fig:circuit-vk} to the reference state. The specific form of $\rho_{k,i}^\phi$ is given by
\begin{align}
	\rho_{k,i}^{\phi} &= 	|\psi^{\phi}_{k,i}\rangle\langle \psi^{\phi}_{k,i}| \\
	|\psi^{\phi}_{k,i}\rangle &= \frac{1}{\sqrt{2}}(|0\rangle R_{k,i}|v_{\rm ref}\rangle + |1\rangle R|v_{\rm ref}\rangle).
\end{align}
The output of the circuit shown in Fig.~\ref{fig:circuit-vk} has the following form,
\begin{align}
\real\left[
	e^{i\phi}
	\bra{v_{\rm ref}}R_{k,i}^{\dagger}P_r R \ket{v_{\rm ref}}
	\right] &=  
	{\rm Tr}\left[(X \otimes I_n) ({\rm CP}_r) \rho^{\phi}_{k,i} ({\rm CP}_r^{\dagger}) \right] \\
	&= {\rm Tr}\left(O_r \rho^{\phi}_{k,i} \right).
\end{align}

In general, given an $n$-qubit unitary operator as $U$, estimating an observable written by $\tilde{O} \equiv U (X\otimes I_n) U^{\dagger}$ requires many two-qubit gates. In fact, previous literature utilizes many two-qubit gates to estimate $O_q$ in \eqref{eq:oq} as we see in Fig.~\ref{fig:circuit-mkl} and Fig.~\ref{fig:circuit-vk}. 
However, in our case, we can write $O_q$ in \eqref{eq:oq} by only one Pauli observable. That is because ${\rm CP}_q $ is a Clifford operator, and $X \otimes I_n$ is a Pauli observable; a Clifford operator generally maps one Pauli observable to another Pauli observable. Direct calculation gives the explicit form of $O_q$ as 
\begin{align}
	O_q = {\rm CP}^{\dagger}_q (X \otimes I_n) {\rm CP}_q &= (|0\rangle\langle 0| \otimes I_n + |1\rangle\langle 1| \otimes P_q)
			(X \otimes I_n)(|0\rangle\langle 0| \otimes I_n + |1\rangle\langle 1| \otimes P_q) \nonumber\\
			&= (|0\rangle\langle 1| \otimes I_n + |1\rangle\langle 0| \otimes P_q)
				(|0\rangle\langle 0| \otimes I_n + |1\rangle\langle 1| \otimes P_q) \nonumber\\
			&= |0\rangle\langle 1| \otimes P_q + |1\rangle\langle 0| \otimes P_q \nonumber\\
			&= X \otimes P_q.
\end{align}
To summarize, we do not need the two-qubit operations in Fig.~\ref{fig:circuit-mkl} and Fig.~\ref{fig:circuit-vk} and we 
can compute each summand of \eqref{eq:mkl-real} and \eqref{eq:vk-real} as 
\begin{align}	
\real\left[
	e^{i\phi} \bra{v_{\rm ref}}R_{k,i}^{\dagger}P_q R_{\ell,i^{\prime}} \ket{v_{\rm ref}}\right] &= 	{\rm Tr}\left[(X \otimes P_q) \rho^{\phi}_{k,i;\ell,i^{\prime}} \right], \\
	\real\left[
	e^{i\phi}
	\bra{v_{\rm ref}}R_{k,i}^{\dagger}P_r R \ket{v_{\rm ref}}
	\right] &= {\rm Tr}\left[(X \otimes P_r) \rho^{\phi}_{k,i} \right].
\end{align}

\subsection{VQS with shadow-based strategies}
\label{section:algorithm-main}
The benefit of writing each $O_q$ and $O_r$ as a Pauli observable is not limited to the reduction of the two-qubit gates.
Although we consider a more general case later, we begin by describing a simplified scenario
that each phase of the summand in \eqref{eq:mkl-real} does not depend on the index $q$ and that in \eqref{eq:vk-real} does not depend on the index $r$; namely 
\begin{align}	
	g_{k,i}^{\ast} g_{\ell, i^{\prime}}\beta_q &= \gm e^{i\phi_{k,i,\ell,i^{\prime}}}~~(\forall q), \\
	g_{k,i}^{\ast}\alpha_r &= \gv  e^{i\phi_{k,i}} ~~(\forall r),
\end{align}
where $\gm$ and $\gv $ are real values, and 
$e^{i\phi_{k,i,\ell,i^{\prime}}}$ and $e^{i\phi_{k,i}}$ are phases independent of indices $q$ and $r$. 
Then, we obtain
\begin{equation}	
	M_{k\ell} = \sum_{i,i^{\prime}}^{\numderiv} M_{k\ell}^{i,i^{\prime}},~ 
	V_k = \sum_{i=1}^{\numderiv}V_k^i,
\end{equation}	
with $M_{k\ell}^{i,i^{\prime}}\equiv \sum_{q=1}^{\nbb} 
	\gm	{\rm Tr}\left[(X \otimes P_q) \rhom \right]$ and $V_k^i=\sum_{r=1}^{\nba} \gv {\rm Tr}\left[(X\otimes P_r)\rho_{k,i}^{\phi_{k,i}} \right]$. Note that $\rhom$ is the density operator $\rho^{\phi}_{k,i;\ell,i^{\prime}}$ with $\phi = \phi_{k,i,\ell,i^{\prime}}$ and $\rhov$ is the density operator $\rho^{\phi}_{k,i}$ with $\phi = \phi_{k,i}$.
		
The notable point here is that $M_{k\ell}^{i,i^{\prime}}$ and $V_k^i$ have the form of 
$\langle \mathcal{H} \rangle =\sum_{j=1}^K a_j {\rm Tr}(P_{j}\rho)$
with $\{a_j\}$ as real coefficients, $\{P_j\}$ as Pauli observables, and $\rho$ as a quantum state. Therefore, we can use shadow-based strategies to efficiently compute the summation inside the brackets. 
Namely, by our estimating $M_{k\ell}^{i,i^{\prime}}$ and $V_k^{i}$ as
\begin{equation}
\begin{split}	
	\nu_{k,i;\ell,i^{\prime}} &= {\bf estimateNu}(\rhom, \{\gm\}_{q=1}^{\nbb}, \{X \otimes P_q\}_{q=1}^{\nbb}, \ns),\\
	\nu_{k,i} &=  {\bf estimateNu}(\rhov, \{G_{k,i,q}\}_{q=1}^{\nba}, \{X\otimes P_q\}_{q=1}^{\nba}, \ns),
\end{split}
\end{equation}
the following values 
\begin{equation}
	\tilde{M}_{k\ell} = \sum_{i,i^{\prime}}^{N_g} \nu_{k,i;\ell,i^{\prime}},
	~\tilde{V}_k = \sum_{i=1}^{N_g}\nu_{k,i},
\end{equation}
become unbiased estimator of $M_{k\ell}$ and $V_k$. We summarize our proposed estimation process of $M_{k\ell}$ and $V_k$ in Algorithm \ref{alg:est_mkl_shadow} and Algorithm \ref{alg:est_vk_shadow}. 
It should be noted that
our circuit synthesis is also useful for
grouping strategies \cite{mcclean2016theory,Verteletskyi2020,Hadfield2022,gokhale2019minimizing,izmaylov2019unitary,crawford2021efficient,huggins2021efficient,hamamura2020efficient,bravyi2017tapering,zhao2020measurement,yen2020measuring,jena2019pauli,yen2022deterministic,choi2022improving,choi2023fluid}, mainly targeting VQO problems, 
where we need to calculate
$M_{k\ell}^{i,i^{\prime}}$ and $V_{k}^{i}$.

The variances of $\tilde{M}_{k\ell}$ and $\tilde{V}_k$ are given by
\begin{equation}
	\var(\tilde{M}_{k\ell}) = \sum_{i,i^{\prime}}^{\numderiv}\var\left(\nu_{k,i;\ell,i^{\prime}}\right),~\var(\tilde{V}_k) = \sum_{i=1}^{\numderiv}\var(\nu_{k,i}).
\end{equation} 
By using \eqref{eq:var} with $\{a_j\}_{j=1}^K=\{\gm\}_{q=1}^{\nbb}$ and $\rho = \rhom$, we obtain 
\begin{equation}
\begin{split}	
\label{eq:shadow-based-variance-m}
	\var(\tilde{M}_{k\ell}) &= \frac{(\Delta M_{k\ell})^2}{N_{\rm shot}}, \\
	\Delta M_{k\ell} &\equiv   \sqrt{\sum_{i,i^{\prime}}^{\numderiv}\left(
		 \sum_{q=1}^{\nbb} \frac{(\gm)^2}{\qfunc{P_q}} + \sum_{q = 1}^{\nbb}\sum_{q^{\prime}\neq q} \gm G_{k,i,\ell,i^{\prime},q^{\prime}} g(\tilde{P}_q, \tilde{P}_{q^{\prime}}){\rm Tr}(\tilde{P}_q \tilde{P}_{q^{\prime}} \rhom)\right) 
		 - M_{k\ell}^2}, 
\end{split}
\end{equation}
where $\tilde{P}_q = X \otimes P_q$,
and by using \eqref{eq:var} with $\{a_j\}_{j=1}^K = \{\gv \}_{q=1}^{\nba}$ and $\rho = \rhov$, we obtain
\begin{equation}
\label{eq:shadow-based-variance-v}
\begin{split}	
	\var(\tilde{V}_{k}) &=  \frac{(\Delta V_k)^2}{\ns}, \\ 
	\Delta V_k &\equiv
	\sqrt{
	\sum_{i}^{\numderiv} 
		\left( \sum_{r=1}^{\nba} \frac{(G_{k,i,r})^2}{\qfunc{P_r}} + \sum_{r = 1}^{\nba}\sum_{r^{\prime}\neq r} \gv  \gvprime \gpp{\rm Tr}(\tilde{P}_r \tilde{P}_{r^{\prime}}  \rhov)\right) - V_k^2
}, 
\end{split}
\end{equation}
where $\tilde{P}_r = X \otimes P_r$. 

The above computed variances when using the shadow-based strategy are different from the ones when we naively estimate each summand of $M_{k\ell}^{i,i^{\prime}}\equiv \sum_{q=1}^{\nbb} 
	\gm	{\rm Tr}\left[(X \otimes P_q) \rhom \right]$ and $V_k^i=\sum_{r=1}^{\nba} \gv {\rm Tr}\left[(X\otimes P_r)\rho_{k,i}^{\phi_{k,i}} \right]$, which is done by the previous literature (we simply call the strategy as the naive strategy in the rest of the paper). 
Let $\numnaive$ be the number of measurements for each circuit in the naive strategy, and let the corresponding estimators of $M_{k\ell}$ and $V_k$ be $\tilde{M}_{k\ell}^{\rm naive}$ and $\tilde{V}_k^{\rm naive}$. Then the variances of $\tilde{M}_{k\ell}^{\rm naive}$ and $\tilde{V}_k^{\rm naive}$ are given by
\begin{equation}
\label{eq:naive-variance}
\begin{split}	
	{\rm Var}(\tilde{M}_{k\ell}^{\rm naive}) &= \frac{(\Delta M_{k\ell}^{\rm naive})^2}{\numnaive},~{\rm Var}(\tilde{V}_k^{\rm naive}) = \frac{(\Delta V_{k}^{\rm naive})^2}{\numnaive},\\
	\Delta M_{k\ell}^{\rm naive} &\equiv \sqrt{\sum_{i,i^{\prime}}^{\numderiv}\sum_{q=1}^{\nbb}\left[(\gm)^2 - \left(\gm{\rm Tr}\left(\tilde{P}_q\rhom\right)\right)^2\right]},
 \\
\Delta V_k^{\rm naive} &\equiv \sqrt{\sum_{i}^{\numderiv} \sum_{r=1}^{\nba}
\left[
(\gv )^2 - 
\left(
\gv {\rm Tr}\left(\tilde{P}_r  \rhov\right)
\right)^2
\right] }.
\end{split}
\end{equation}
The difference in the variances between the naive strategy and the shadow-based strategy results in the difference in the infidelity due to the shot noise for a given number of measurements. In the following subsection, we quantitatively evaluate how the variance affects the infidelity and clarifies the benefit of using the shadow-based strategy in the VQS.

 Note that in the above introduction of our proposed algorithm, we require that each phase of the summand in \eqref{eq:mkl-real} does not depend on the index $q$ and that in \eqref{eq:vk-real} does not depend on the index $r$. 
 However, we can remove the requirement as follows. Instead of \eqref{eq:bb} , we can expand $B^{\dagger}(t)B(t)$  as 
\begin{align}	
	B^{\dagger}(t)B(t) &= \sum_{q=1}^{\nbb}\real(\beta_q) P_q +  \sum_{q=1}^{\nbb} \imaginary (\beta_q) P_q, 
\end{align}
where the symbol $\imaginary(z)$ is the imaginary part of $z$. Then, we can  evaluate $\sum_{q=1}^{\nbb}\real(\beta_q) P_q$ and $ \sum_{q=1}^{\nbb} \imaginary (\beta_q) P_q$ independently by the shadow-based strategy though the number of measurements increase. That is also true for the computation of $B^{\dagger}(t)A(t)$. 

\subsubsection*{Hybrid strategy in RTE and ITE}

In RTE and ITE, 
$\nbb = 1$ is set to calculate $M_{k\ell}$
 and there are no advantages for using a shadow-based strategy for the computation of $M_{k\ell}$. Therefore, in RTE and ITE, we take the hybrid strategy, meaning that we estimate $V_k$ by the shadow-based strategy and $M_{k\ell}$ by the naive strategy. At the end of Section~\ref{section:algorithm-advantage}, we discuss the advantage of the hybrid strategy. 

\begin{algorithm}[ht]
  \caption{Estimation of $M_{k\ell}$ with a shadow-based strategy.}\label{alg:est_mkl_shadow}
  \begin{algorithmic}[1]
  \INPUT $\rhom$, $\{\gm\}_{q=1}^{\nbb}$, $\{P_q\}_{q=1}^{\nbb}$, $\ns$, $\numderiv$
     \STATE SET result~$= 0$.
  	\FOR{$i=1...\numderiv$} 
	  	\FOR{$i^{\prime}=1...\numderiv$}
	  	\STATE SET $\nu_{k,i;\ell,i^{\prime}} = $ {\bf estimateNu}($\rhom$, $\{\gm\}_{q=1}^{\nbb}$, $\{X \otimes P_q\}_{q=1}^{\nbb}$, $\ns$).
	  	\STATE SET result = result +  $\nu_{k,i;\ell,i^{\prime}}$.
  	  \ENDFOR
  	\ENDFOR
    \RETURN result
  \end{algorithmic}
\end{algorithm}

\begin{algorithm}[ht]
  \caption{Estimation of $V_{k}$ with a shadow-based strategy.}\label{alg:est_vk_shadow}
  \begin{algorithmic}[1]
  \INPUT $\rhov$, $\{G_{k,i,q}\}_{q=1}^{\nba}$, $\{P_q\}_{q=1}^{\nba}$, $\ns$, $\numderiv$
     \STATE SET result~$= 0$.
  	\FOR{$i=1...\numderiv$} 
	  	\FOR{$i^{\prime}=1...\numderiv$}
	  	\STATE SET $\nu_{k,i} = $ {\bf estimateNu}($\rhov$, $\{G_{k,i,q}\}_{q=1}^{\nba}$, $\{X\otimes P_q\}_{q=1}^{\nba}$, $\ns$).
	  	\STATE SET result = result +  $\nu_{k,i}$.
  	  \ENDFOR
  	\ENDFOR
    \RETURN result
  \end{algorithmic}
\end{algorithm}

\subsection{Advantage of using shadow-based strategy}
\label{section:algorithm-advantage}
To clarify the advantage of using the shadow-based strategy over the naive strategy, we compare the infidelity due to the shot noise after the duration $T$.  
The time derivative of the parameter vector can be ideally computed as 
\begin{equation}
\label{eq:ideal-time-evolution}
	\deriv{\vec{\theta}_{\rm ideal}(t)}{t} = M^{-1} \vec{V}.
\end{equation}
However, we estimate $M$ and $\vec{V}$ with a finite number of measurements, and therefore, the estimated time derivative includes the shot noise: 
\begin{equation}
\label{eq:distored-time-evolution}
	\deriv{\vec{\theta}(t)}{t} = (M + \delta M)^{-1} (\vec{V} + \delta \vec{V}),
\end{equation} 
where $\delta M$ and $\delta \vec{V}$ are the error to $M$ and $\vec{V}$ 
due to
the shot noise. As in literature \cite{Li2017}, we use the following infidelity as a metric for evaluating the effect of error: 
\begin{equation}
\label{eq:distance}
\di = \sqrt{1 - 	\left|\left\langle v(\vecthetaidealT)\right| \left. v(\vecthetaT)\right\rangle\right|^2},
\end{equation}
 between $\ket{v(\vecthetaidealT)}$ evolved by the ideal time derivative \eqref{eq:ideal-time-evolution} and $\ket{v(\vecthetaT)}$ evolved by the distorted time derivative \eqref{eq:distored-time-evolution} at time $T$.

 Suppose that $D_I^{\rm naive}$ is an upper bound of $D_I$ when using the naive strategy and $D_I^{\rm shadow}$ is the one when using the shadow-based strategy. 
 As we show in \eqref{eq:num-measurement-ratio} in Appendix~\ref{section:shot-noise}, $D_I^{\rm naive}$ = $D_I^{\rm shadow}$ holds when
\begin{equation}
\label{eq:num-measurement-ratio-maintext}
	\frac{N_{\rm shot}}{\numnaive} \simeq \left(\frac{\Deltashadow^{\rm max}}{\Deltanaive^{\rm max}}\right)^2,  
\end{equation}
where 
\begin{equation}
	\begin{split}
				\Deltanaive &= || M^{-1} ||_F \sqrt{\sum_k (\Delta V_k^{\rm naive})^2} 
	+ ||M^{-1}||_F^2 ||\vec{V}||_2 \sqrt{\sum_{k\ell}(\Delta M_{k\ell}^{\rm naive})^2}, \\
	\Deltashadow &= || M^{-1} ||_F \sqrt{\sum_k (\Delta V_k)^2} 
	+ ||M^{-1}||_F^2 ||\vec{V}||_2 \sqrt{\sum_{k\ell}(\Delta M_{k\ell})^2}, 
	\end{split}
\end{equation}
$\Deltanaive^{\rm max} \equiv \max_{0\leq t \leq T}( \Deltanaive)$, 
$\Deltashadow^{\rm max} \equiv \max_{0\leq t \leq T}( \Deltashadow)$, $||\cdot ||_F$ is the Frobenius norm, and $||\cdot||_2$ is the two-norm. 

Note that the total number of measurements for each step in the case of the shadow-based strategy is $N_{\rm shadow}^{\rm total} = \ns (\numparams \numderiv + \numparams^2\numderiv^2 )$ while that in the naive strategy is $\numnaive^{\rm total} = \numnaive(\numparams \numderiv \nba + \numparams^2 \numderiv^2 \nbb)$. 
Thus, the advantage of using a shadow-based strategy arises if $N^{\rm total}_{\rm shadow}/N^{\rm total}_{\rm naive} < 1$ when \eqref{eq:num-measurement-ratio-maintext} is satisfied, i.e., 
\begin{equation}
\label{eq:mainRatio}
\frac{N_{\rm shadow}^{\rm total}}{N_{\rm naive}^{\rm total}} = \frac{1 + \numparams \numderiv}{\nba + \numparams \numderiv \nbb} 
 \left(\frac{\Deltashadow^{\rm max}}{\Deltanaive^{\rm max}}\right)^2 < 1.
\end{equation}
In the rest of this section, we give a rough estimate of the leftest-hand side of the inequality. 

\subsubsection*{Estimation of ${N_{\rm shadow}^{\rm total}}/{N_{\rm naive}^{\rm total}} $}
We assume that the values of $||M^{-1}||_F$ and $||M||_F^{2} ||\vec{V}||_2$ when using the naive strategy is almost the same as those when using the shadow-based strategy. The assumption is valid since the time evolutions with both strategies are close to each other (recall that we consider the regime where $D_I^{\rm naive} = D_I^{\rm shadow}$). Under the assumption, we obtain the inequality: 
\begin{equation}
\label{eq:shadow-naive-ratio-bound}
\frac{\Deltashadow^{\rm max}}{\Deltanaive^{\rm max}} <
	\sqrt{\frac{\sum_k (\Delta V_k)^2}{\sum_k (\Delta V_k^{\rm naive})^2}} + 
	\sqrt{\frac{\sum_{k\ell} (\Delta M_{k\ell})^2}{\sum_{k\ell} (\Delta M_{k\ell}^{\rm naive})^2}}. 
\end{equation}
We evaluate each term on the right-hand side in the following. 

For the numerator in the square root of the first term, it holds 
\begin{equation}
	(\Delta V_k)^2 \leq \sum_{i=1}^{\numderiv} 
		\vki (\rhov),
\end{equation}
where 
\begin{equation}
\label{eq:vkrho}
			\vki (\rho) = \sum_{r=1}^{\nba} \frac{(\gv)^2}{\qfunc{P_r}} + \sum_{r = 1}^{\nba}\sum_{r^{\prime}\neq r} \gv  \gvprime \gpp{\rm Tr}(\tilde{P}_r \tilde{P}_{r^{\prime}}  \rho).
\end{equation}
Since $\vki (\rho)$ depends on the quantum state $\rho$, it is difficult to directly estimate the value. Instead of evaluating $\vki (\rho)$, let us estimate the typical size of $\vki (\rho)$ in the overall Hilbert space by computing 
\begin{equation}
\label{eq:haar-averagev}
	\langle \vki  \rangle_{\rm Haar} = \int_{\rm Haar} dU 
	\vki (U|0\rangle\langle 0|U^{\dagger}),~{\rm and}~
		{\rm Var} (\vki )_{\rm Haar} = \int_{\rm Haar} dU 
	\left(\vki (U|0\rangle\langle 0|U^{\dagger})\right)^2 - (\langle \vki  \rangle_{\rm Haar})^2,
\end{equation}
where $\int_{\rm Haar}$ is the average over the Haar distribution. 
By calculating those statistical values, we can see how $\vki(\rho)$ distributes when $\rho$ is uniformly sampled. 
As we show in Appendix~\ref{section:haar-integral}, we can readily show 
\begin{equation}
\label{eq:vk-haar-integral}
\begin{split}	
	\langle \vki  \rangle_{\rm Haar} &= 	 \sum_{r=1}^{\nba} \frac{(\gv)^2}{\qfunc{P_r}}, \\
	{\rm Var} (\vki )_{\rm Haar} &=  \frac{1}{2^{n+1}(2^{n+1} + 1)} 
\sumru	 \gv  \gvprime 	 \gvu \gvuprime 
	 \gpp \gppu
	 {\rm Tr}(\tilde{P}_r \tilde{P}_{r^{\prime}} \tilde{P}_u \tilde{P}_{u^{\prime}}),
\end{split}
\end{equation}
where $n$ is the number of qubits in PQC. 
By using ${\rm Tr}(\tilde{P}_r \tilde{P}_{r^{\prime}} \tilde{P}_u \tilde{P}_{u^{\prime}}) \leq 2^{n+1}$, we can show
$ {\rm Var}  (\vki )  \propto O(1/2^{n})$, which 
becomes asymptotically zero as $n$ becomes large. 
	Since the variance is exponentially suppressed, $\vki (\rhov)$ takes the value almost equal to the average with high probability when uniformly choosing $\rhov$.
	In the rest of this section, given a function of a density operator as $\xi(\rho)$, if $\langle \xi(\rho) \rangle_{\rm Haar} = \xi_0$ holds with $\xi_0$ as a constant and ${\rm Var}(\xi)_{\rm Haar}$ is exponentially suppressed, we denote $\xi(\rho) \simeq \xi_0$. Also, for given another function of a density operator as $\xi^{\prime}(\rho)$, if $\xi^{\prime}(\rho) \leq \xi(\rho) \simeq \xi_0$, we denote $\xi^{\prime}(\rho) \lesssim \xi_0$. 
Thus, with a large $n$,
\begin{equation}	
\label{eq:delv-bound}
\vki (\rhov) \simeq \sum_{r=1}^{\nba}  \frac{(\gv)^2}{\qfunc{P_r}}
\end{equation}
holds and 
\begin{equation}
		(\Delta V_k)^2 \lesssim \sum_{i=1}^{\numderiv}  \sum_{r=1}^{\nba}  \frac{(\gv)^2}{\qfunc{P_r}}.
\end{equation}

In the Eq. \eqref{eq:shadow-naive-ratio-bound}, for the denominator of the square root in the first term, 
\begin{equation}
	(\Delta V_k^{\rm naive})^2 = \sum_{i=1}^{\numderiv} \sum_{r=1}^{\nba} 
	\gkir( \rhov),
\end{equation}
where
\begin{equation}
	\gkir(\rho) = (\gv )^2 \left[ 1 - 
\left({\rm Tr}(\tilde{P}_r \rho)\right)^2
\right].  
\end{equation}
As in the case of $\vki (\rho)$, we can compute the average in the overall Hilbert space (See Appendix~\ref{section:haar-integral})
as 
\begin{equation}
\label{eq:vknaive-haar-integral}
\begin{split}	
	\langle \gkir(\rho) \rangle_{\rm Haar} &=  (\gv)^2 \left(1 - \frac{1}{2^{n+1} + 1} \right),\\
	{\rm Var} \left(\gkir(\rho)\right)_{\rm Haar} &= 
	\frac{ (\gv)^4 (2\cdot 2^{2(n+1)} + 8\cdot 2^{2(n+1)})}{(2^{n+1} + 1)^2(2^{n+1}+3)}.
\end{split}
\end{equation}
Thus, for a large enough $n$,
$\gkir(\rho)\simeq (\gv)^2$
and, we obtain 
\begin{equation}
\label{eq:delv-naive-bound}
	\left(\Delta V_k^{\rm naive}\right)^2 \simeq \sum_{i=1}^{\numderiv} \sum _{r=1}^{\nba}(\gv)^2.
\end{equation}

By combining \eqref{eq:delv-bound} and \eqref{eq:delv-naive-bound}, the first term in the right hand side of \eqref{eq:shadow-naive-ratio-bound} is bounded as 
\begin{equation}
\label{eq:first-term-bound}
	\sqrt{\frac{\sum_k (\Delta V_k)^2}{\sum_k (\Delta V_k^{\rm naive})^2}} \lesssim  
	\sqrt{\left\langle\frac{1}{q(\tilde{P}_r)}\right\rangle_{\gv}}, 
\end{equation}
 where $\langle \cdot \rangle_{\gv}$ is a weighted average defined by 
 \begin{equation}
 	\left\langle \mathcal{F}_r \right\rangle_G = \frac{
	\sum_{k=1}^{\numparams} \sum_{i=1}^{\numderiv}  \sum_{r=1}^{\nba}  (\gv)^2 \mathcal{F}_r}{\sum_{k=1}^{\numparams} \sum_{i=1}^{\numderiv}  \sum_{r=1}^{\nba} (\gv)^2}.
 \end{equation}
 Similarly, 
 \begin{equation}
 \label{eq:second-term-bound}
 	\sqrt{\frac{\sum_{k\ell} (\Delta M_{k\ell})^2}{\sum_{k\ell} (\Delta M_{k\ell}^{\rm naive})^2}} \lesssim 
 	\sqrt{\left\langle \frac{1}{q(\tilde{P}_q)}\right\rangle_{\gm}}, 
 \end{equation}
 where $\langle \cdot \rangle_{\gv}$ is another weighted average defined by 
 \begin{equation}
 	\left\langle \mathcal{F}_q \right\rangle_{\gm} = \frac{\sum_{k,\ell}^{\numparams} \sum_{i,i^{\prime}}^{\numderiv}  
 	\sum_{q=1}^{\nbb}  (\gm)^2 \mathcal{F}_q}{\sum_{k,\ell}^{\numparams} \sum_{i,i^{\prime}}^{\numderiv}  \sum_{q=1}^{\nbb} (\gm)^2}. 	
 \end{equation}
By substituting \eqref{eq:first-term-bound} and \eqref{eq:second-term-bound} to 
\eqref{eq:shadow-naive-ratio-bound}, we obtain
\begin{equation}
\label{eq:total-ratio}
\frac{\Deltashadow^{\rm max}}{\Deltanaive^{\rm max}} \lesssim 	\sqrt{\left\langle\frac{1}{q(\tilde{P}_r)}\right\rangle_{\gv}} +  	\sqrt{\left\langle \frac{1}{q(\tilde{P}_q)}\right\rangle_{\gm}}. 
\end{equation}
Combining with \eqref{eq:mainRatio}, we obtain
\begin{equation}
\frac{N_{\rm shadow}^{\rm total}}{\numnaive^{\rm total}} \lesssim \frac{1 + \numparams \numderiv}{\nba + \numparams \numderiv \nbb} 
 \left(
\sqrt{\left\langle\frac{1}{q(\tilde{P}_r)}\right\rangle_{\gv}} +  	\sqrt{\left\langle \frac{1}{q(\tilde{P}_q)}\right\rangle_{\gm}}
 \right)^2.  
\end{equation}

The factor ${(1 + \numparams \numderiv})/({N_{BA} + \numparams \numderiv N_{BB})} < 1$ corresponds to the benefit of simaltaneously evaluating the $\nba$ terms in the calcualtion of $V_k$ and the $\nbb$ terms in the calculation of $M_{k\ell}$ in the shadow-based strategy. The other factor $\sqrt{\left\langle{1}/{q(\tilde{P}_r)}\right\rangle_{\gv}} + \sqrt{\left\langle {1}/{q(\tilde{P}_q)}\right\rangle_{\gm}}$ corresponds to how well measurements in the shadow-based strategy cover the observables; if observables are not covered efficiently, the benefit of the simaltaneous evaluation may disappear due to this factor. Particularly, when we use the classical shadow as the shadow-based strategy, we can explicitly write the latter factor as  
\begin{equation}
 \left(
\sqrt{\left\langle\frac{1}{q(\tilde{P}_r)}\right\rangle_{\gv}} +  	\sqrt{\left\langle \frac{1}{q(\tilde{P}_q)}\right\rangle_{\gm}}
 \right)^2 = \left(
\sqrt{\left\langle 3^{\rm locality}(\tilde{P}_r) \right\rangle_{\gv}} +  	\sqrt{\left\langle3^{\rm locality}(\tilde{P}_q) \right\rangle_{\gm}}
 \right)^2,   	
\end{equation}
where we use \eqref{eq:covering-cs}. Thus, the locality of the observables should be small in order for enjoying the benefit of the classical shadow
in the VQS. It should be noted that 
\begin{equation}	
\label{eq:locality-increase}
{\rm locality}(\tilde{P}_r) = {\rm locality}(P_r)+ 1, 
\end{equation}
because $\tilde{P}_r = X \otimes P_r$.

\subsubsection*{Advantages of using the shadow-based strategy in RTE and ITE}

Finally, we discuss the advantage of using the shadow-based strategy in RTE and ITE. As we noted at the end of Section~\ref{section:algorithm-main}, we take the hybrid strategy that utilizes the shadow-based strategy for computing $V_k$ and the naive strategy for computing $M_{k\ell}$. 
Unlike the general case (where we measure both $V_k$ and $M_{k\ell}$ by using the shadow-based strategy), we can reduce the number of measurements for estimating each $M_{k\ell}$ to smaller than that for estimating $V_k$ (=$\ns$); we set the number of measurements as $\ns \alpha$ with $0 < \alpha < 1$. 

Let 
\begin{equation}
	\Deltahybrid = || M^{-1} ||_F \sqrt{\sum_k (\Delta V_k)^2} 
	+ ||M^{-1}||_F^2 ||\vec{V}||_2 \sqrt{\frac{\sum_{k\ell}(\Delta M_{k\ell}^{\rm naive})^2}{\alpha}}. 
\end{equation} 
Then, as shown in \eqref{eq:num-measurement-ratio-hybrid} in Appendix~\ref{section:shot-noise}, we can show that 
\begin{equation}
\label{eq:num-measurement-ratio-hybrid-maintext}
	\frac{N_{\rm shot}}{\numnaive} \simeq \left(\frac{\Deltahybrid^{\rm max}}{\Deltanaive^{\rm max}}\right)^2,  
\end{equation}
if the error due to the shot noise in the naive strategy is almost the same as that in the hybrid strategy,   
where we define $\Deltahybrid^{\rm max} \equiv \max_{0\leq t \leq T}( \Deltahybrid)$.
Given $N_{\rm hybrid}^{\rm total}$ as the total number of measurements for each step in the hybrid strategy, it holds $N_{\rm hybrid}^{\rm total} = \ns (\numparams \numderiv + \alpha(\numparams \numderiv)^2)$. Under the same assumption to obtain \eqref{eq:shadow-naive-ratio-bound}, it holds
\begin{equation}
\label{eq:ratio-hybrid}
\frac{N_{\rm hybrid}^{\rm total}}{\numnaive^{\rm total}} \lesssim \frac{1 + \numparams \numderiv\alpha}{\nba + \numparams \numderiv } 
 \left(
\sqrt{\left\langle\frac{1}{q(\tilde{P}_r)}\right\rangle_{\gv}} + \sqrt{\frac{1}{\alpha}}\right)^2 < \frac{1 + \numparams \numderiv \alpha}{\nba} \left(
\sqrt{\left\langle\frac{1}{q(\tilde{P}_r)}\right\rangle_{\gv}} + \sqrt{\frac{1}{\alpha}}\right)^2.
\end{equation}
By setting $\alpha = 4/\left\langle\frac{1}{q(\tilde{P}_r)}\right\rangle_{\gv} $, 
\begin{equation}
	\frac{N_{\rm hybrid}^{\rm total}}{\numnaive^{\rm total}} 
	\lesssim \frac{1 + N_\alpha}{\nba} \times\left\langle\frac{1}{q(\tilde{P}_r)}\right\rangle_{\gv} \times \frac{9}{4},
\end{equation}
with 
\begin{equation}
	N_{\alpha} = N_p N_g \alpha = \frac{4\numparams\numderiv}{\left\langle\frac{1}{q(\tilde{P}_r)}\right\rangle_{\gv}}.
\end{equation}

Note that the literature \cite{van2021measurement} states that given a circuit structure, the number of measurements for estimating $\{M_{k\ell}\}_{k,\ell}^{\numparams}$ asymptotically becomes negligible compared to that for estimating $\{V_k\}_{k=1}^{\nba}$ as the number of qubits $n$ increases and $\nba$ increases faster than quadratically with $n$. Though the literature assumes the naive strategy to estimate $\{V_k\}_{k=1}^{\nba}$, the asymptotic behavior would be the same in our hybrid strategy. 
The condition that the number of measurements for $\{M_{k\ell}\}_{k,\ell}^{\numparams}$ is negligible, corresponds to the case when $N_{\alpha} \ll 1$. In that case, it holds
\begin{equation}
\label{eq:condition-for-shadow-advantage}
	\frac{N_{\rm hybrid}^{\rm total}}{\numnaive^{\rm total}} 
	\lesssim \frac{1}{\nba } \times \left\langle\frac{1}{q(\tilde{P}_r)}\right\rangle_{\gv} \times \frac{9}{4}.
\end{equation}

For example, let us calculate the right-hand side value of the right-hand side of \eqref{eq:condition-for-shadow-advantage} when the time evolution Hamiltonian is given by \eqref{eq:easyHamiltonian}. We can write
\begin{equation}
\label{eq:extendedEasyHamiltonian}
	X \otimes H = XXXXZ + XXXII +   XIIXZ + XYYZX + XYYII + XIIZX,
\end{equation}
and therefore, the number of terms is given by $\nba = 6$. Recall that each $\tilde{P}_r$ in \eqref{eq:condition-for-shadow-advantage} corresponds to each term in \eqref{eq:extendedEasyHamiltonian}. 
Let us first calculate when we use the classical shadow.
Since the locality of each term in \eqref{eq:extendedEasyHamiltonian} is equal to or larger than three, 
\begin{equation}
	\left\langle\frac{1}{q(\tilde{P}_r)}\right\rangle_{\gv} \geq 3^3, 
\end{equation}
where we use \eqref{eq:covering-cs}.
Thus, in the case of the classical shadow, the right-hand side of \eqref{eq:condition-for-shadow-advantage} is 
\begin{equation}
	 \frac{1}{\nba } \times \left\langle\frac{1}{q(\tilde{P}_r)}\right\rangle_{\gv} \times \frac{9}{4} \geq 10,
\end{equation}
and there is no advantage in using the classical shadow with the Hamiltonian \eqref{eq:easyHamiltonian}. On the contrary, when using derandomization, half of the chosen measurements are $XXXXZ$, and the other half is $XYYZX$ (see the discussion around \eqref{eq:easyHamiltonian}). Therefore, 
\begin{equation}
	\left\langle\frac{1}{q(\tilde{P}_r)}\right\rangle_{\gv} = \frac{1}{2}, 
\end{equation}
and the right-hand side of \eqref{eq:condition-for-shadow-advantage} is 
\begin{equation}
	 \frac{1}{\nba } \times \left\langle\frac{1}{q(\tilde{P}_r)}\right\rangle_{\gv} \times \frac{9}{4} = \frac{3}{4}.
\end{equation}
Thus, there is merit in using derandomization with the Hamiltonian \eqref{eq:easyHamiltonian}.
We note that we utilize the Hamiltonian with just six terms in this example, and the advantage of using the shadow-based methods is limited. In Section~\ref{section:numerical-demonstration}, we will demonstrate the result of VQS when using the molecular Hamiltonian and see the significant advantages of using the shadow-based methods. 

\subsection{Remark on the general shadow-based strategies}
\label{section:remark}
In Section~\ref{section:algorithm-advantage}, we showed that 
\begin{equation}
	(\Delta V_k)^2 \lesssim \sum_{i=1}^{\numderiv} \sum_{r=1}^{\nba}\frac{G_{k,i,r}}{q(P_r)},~{\rm and}~
	(\Delta V_k^{\rm naive})^2 \simeq \sum_{i=1}^{\numderiv} \sum_{r=1}^{\nba}  (G_{k,i,r}^2) \nonumber
\end{equation}
in the context of VQS. 
The same argument holds even in the general estimation problem of $\langle \mathcal{H}\rangle = \sum_{j=1}^K a_j {\rm Tr}(P_j \rho)$ discussed in Section~\ref{section:shadow-based-strategy}. Let $\nu_{\rm naive}$ be the estimator of $\langle \mathcal{H}\rangle$ when estimating each term of $\langle \mathcal{H}\rangle$ one by one with $N_{\rm naive}$ measurements. Also, recall that $\nu$ is the estimator of $\langle \mathcal{H}\rangle$ when using a shadow-based strategy with $\ns$ measurements. Then, with the same arguments in Section~\ref{section:algorithm-advantage}, it can be shown that
\begin{align}	
{\rm Var}(\nu) \lesssim  \frac{1}{\ns}
		 \sum_{j=1}^K \frac{a_j^2}{\qfunc{P_j}}, \\
{\rm Var}(\nu_{\rm naive}) \simeq \frac{1}{N_{\rm naive}}\sum_{j=1}^K a_j^2. 
\end{align}
The total number of measurements is $N_{\rm shadow}^{\rm total} = \ns$ in the case of using the shadow-based strategy and $N_{\rm naive}^{\rm total} = K N_{\rm naive}$ in the case of naively measuring each term one by one. 
For achieving ${\rm Var}(\nu) = {\rm Var}(\nu_{\rm naive})$, the following relation is necessary
\begin{equation}
	\frac{N_{\rm shadow}^{\rm total}}{N_{\rm naive}^{\rm total}} \lesssim \frac{1}{K} 
	\frac{		 \sum_{j=1}^K \frac{a_j^2}{\qfunc{P_j}}}{ \sum_{j=1}^K a_j^2}.
\end{equation}
Notably, as long as the covering probability $\qfunc{P_j}$ is $O(1)$, we can reduce the number of measurements by the factor $O(1/K)$ in the shadow-based strategy.

The above observations consolidate the validity of using the shadow-based strategies \cite{Huang2020,Hadfield2022,Huang2021,Hillmich2021,Wu2021} for estimating the sum of Pauli observables. 
In Eq. (18) of \cite{Hadfield2022}, they also give the upper bound for the variance of $\nu$. However, it is as loose as 
\begin{equation}	
\label{eq:shadow-naive-upper-bound}
 {\rm Var}(\nu) \leq \frac{K}{\ns} \sum_{j=1}^K \frac{a_j^2}{\qfunc{P_j}}, 
 \end{equation}
in our notation (in \cite{Hadfield2022}, they write the number of terms as $|H_0|$ instead of $K$). We can easily check that if ${\rm Var}(\nu)$ takes the value of the upper-bound \eqref{eq:shadow-naive-upper-bound}, $N_{\rm shadow}^{\rm total} > N_{\rm naive}^{\rm total}$, and there are no advantages of using shadow-based strategies. Conversely, previous research does not theoretically show the advantages of using a shadow-based strategy to estimate the sum of Pauli observables. Our upper bound, however, successfully removes the effect from the second term in \eqref{eq:var} by using the average over the whole quantum state and shows that in most of the quantum states $\rho$, there is an advantage to using the shadow-based strategy. 
 


\section{Numerical Demonstration}
\label{section:numerical-demonstration}
Here we demonstrate our shadow-based measurement optimization strategy for the VQS. In Section~\ref{section:our-algorithm}, we generally discussed our strategy, but in this demonstration, we focus on RTE and ITE, which is the algorithm most actively studied in the VQS algorithms. As we noted at the end of Section~\ref{section:algorithm-main}, we take the hybrid strategy, meaning that we obtain $M_{k\ell}$ by the naive strategy and $V_k$ by the shadow-based strategy. 

We perform two types of demonstrations. The first one is the evaluation of the infidelity due to the shot noise, 
\begin{equation}
	\di = \sqrt{1 - 	\left|\left\langle v(\vecthetaidealT)\right| \left. v(\vecthetaT)\right\rangle\right|^2} \nonumber,
\end{equation}
which is defined in \eqref{eq:distance}, in various Hamiltonians, and in various measurement strategies. The second one is the evaluation of the variances in each strategy. Particularly, as a result of the Haar average in the Hilbert space, 
we obtained 
\begin{equation}
		(\Delta V_k)^2 \lesssim \sum_{i=1}^{\numderiv}  \sum_{r=1}^{\nba}  \frac{(\gv)^2}{\qfunc{P_r}},~\left(\Delta V_k^{\rm naive}\right)^2 \simeq \sum_{i=1}^{\numderiv} \sum_{r=1}^{\nba}   (\gv)^2,
\nonumber
\end{equation}
in \eqref{eq:delv-bound} and \eqref{eq:delv-naive-bound} in Section~\ref{section:algorithm-advantage}. We check if it holds in our numerical experiments. 

Before going into each experiment, let us give 
the setting of numerical experiments
in Section~\ref{section:exp-settings}. Then, in the subsequent sections, we demonstrate the results of the numerical
experiments. 

\subsection{Setting of numerical experiments}
\label{section:exp-settings}
Here we summarize Hamiltonians, the reference state, and the ansatz used in the experiments. 
Also, we describe strategies to estimate $M_{k\ell}$ and $V_k$.

\subsubsection*{Hamiltonians}
We use three types of problem Hamiltonians $H$: 
\begin{itemize}
	\item ${\rm H}_2$ molecule in the 6-31g basis with the Bravyi-Kitaev transformation (8-qubit, $\nba=184$);
	\item ${\rm LiH}$ molecule in the STO-3g basis with the Bravyi-Kitaev transformation (12-qubit, $\nba=630$);
	\item Heisenberg model Hamiltonian (6-qubit, $\nba=24$).
\end{itemize} 
We write the number of qubits in the Hamiltonian as $n$.
Here we define our Heisenberg model Hamiltonian by
\begin{equation}
\label{eq:heisenberg-Hamiltonian}
	H = 0.1 \sum_{j=1}^6 \sum_{a=x,y,z} \sigma_a^j \sigma_{a}^{j+1}  + 0.1 \sum_{j=1}^6 \sigma_z^j,
\end{equation}
with $\sigma_a^j$ as a Pauli operator operating on the $j$-th qubit, and $\sigma_a^7 = \sigma_a^1$. 
\subsubsection*{Reference state and ansatz}
As the reference state $\ket{v_{\rm ref}}$, we use the $\ket{v_{\rm ref}} = {\rm H}^{\otimes n} |0\rangle^{\otimes n}$. 
As the choice of the ansatz $R$ (defined in \eqref{eq:ansatz}), we use a hardware efficient ansatz, which contains multiple layers of single-qubit rotation gates and entanglers. Each parameter is embedded in each single-qubit rotation gate as 
$\exp(-i \sigma_a \theta_j/2 )$, where $\sigma_a$ is a Pauli matrix with $a \in \{x, y, z\}$  (therefore, $\numderiv = 1$ and $g_{k,1} = -i/2$). The type of each single-qubit rotation gate `$a$' is randomly chosen at the initialization. 
As the two-qubit gates in entanglers, we use the controlled-Z gates. 
At each layer, we first perform the single qubit gates and, secondly, perform the controlled-Z gates.
The way of operating the controlled-Z gates is different in the odd layers and the even layers. Let us define a set of integers $\mathcal{N}_1$ by $\mathcal{N}_1 = \{m_1| m_1\in {\bf N}~and~1 < 2m_1 \leq n\}$ and another set of integers $\mathcal{N}_2$ by $\mathcal{N}_2 = \{m_2| m_2\in {\bf N}~and~1 < 2m_2 + 1 \leq n\}$. 
For example, when $n=8$, $\mathcal{N}_1 = \{1, 2, 3, 4\}$ and $\mathcal{N}_2 = \{1, 2, 3\}$.
Also, we denote by $(j, k)$ the pair of control and target qubits 
on which we perform a two-qubit operation.
  Then, in each odd layer, pairs of
  $(2m_1-1, 2m_1)$ for all $m_1 \in \mathcal{N}_1$ are firstly entangled by the controlled-Z gates, and next, pairs
of $(2m_2, 2m_2+1)$ for all $m_2 \in \mathcal{N}_2$ are entangled. On the contrary, in each even layer, pairs $(2m_2, 2m_2+1)$ for all $m_2 \in \mathcal{N}_2$ are firstly entangled by the controlled-Z gates, and next, pairs $(2m_1-1, 2m_1)$ for all $m_1 \in \mathcal{N}_1$ are entangled. In Fig.~\ref{fig:hea}, we show our ansatz with the reference state $\ket{v_{\rm ref}}$ when $n=8$ as an example. For all experiments, we fix the number of layers as four, and therefore, $\numparams = 4 n$. All parameters are initialized as zero, and therefore, the initial state generated by the ansatz is $|v_{\rm ref}\rangle$.

  \fig{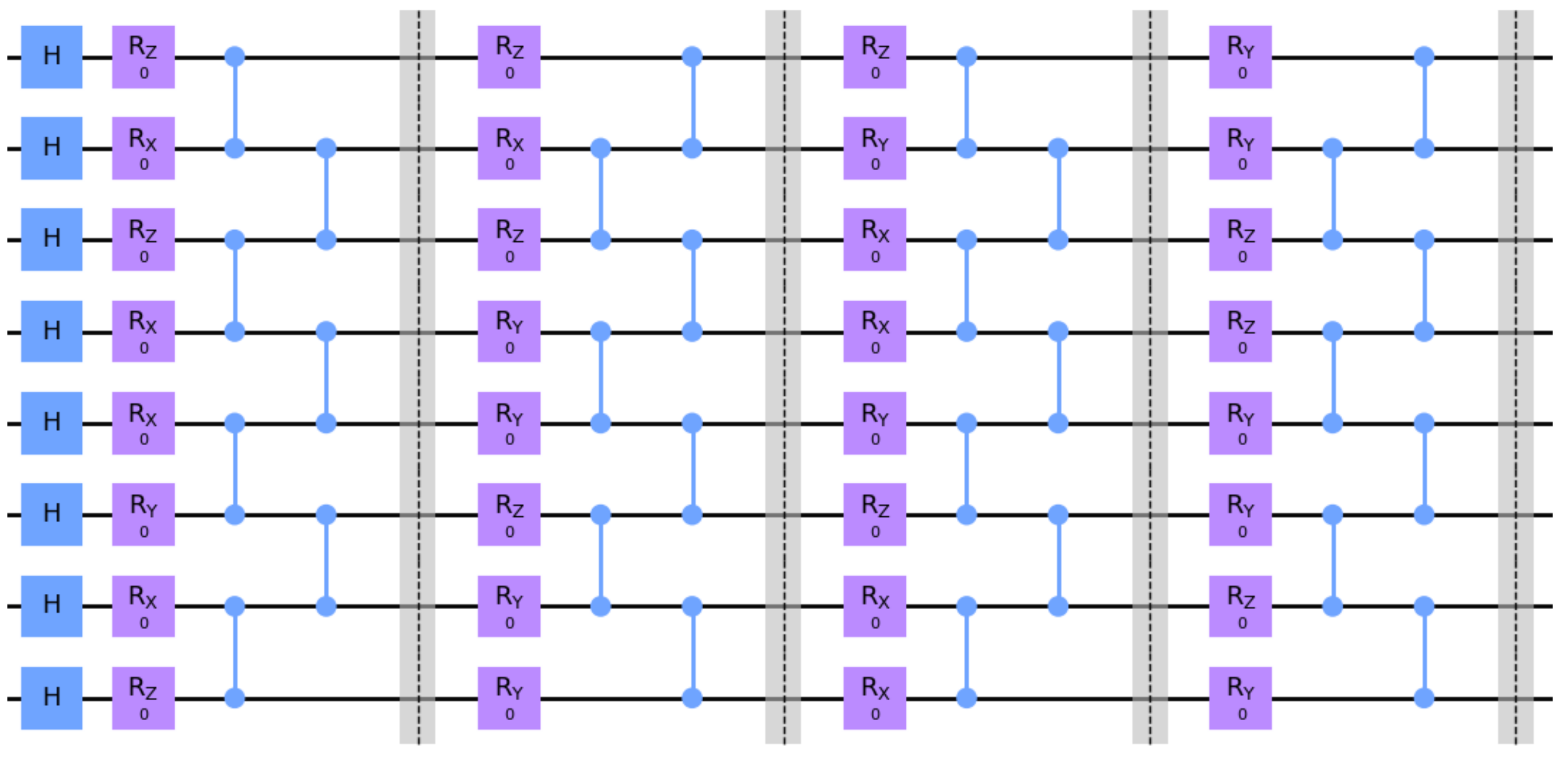}{400pt}{The reference state and the ansatz used in the numerical experiment. As an example, we shot the case when $n=8$.}{fig:hea}
  
\subsubsection*{Strategies to estimate $M_{k\ell}$ and $V_k$}
For the estimation of $V_k$, we compare following four strategies including two shadow-based strategies:
\begin{itemize}
	\item Classical shadow (=a shadow-based strategy);
	\item Derandomization (=another shadow-based strategy);
	\item Naively estimate each term of $V_k$ one by one;
	\item Largest degree first (LDF) grouping \cite{Verteletskyi2020,Hadfield2022}.
\end{itemize}
The LDF grouping is a method to reduce the number of measurements by grouping the simultaneously measurable Pauli observables. In the LDF grouping, the measurement basis is chosen so that all of the observables in one of the groups should be covered by each measurement. 
The group to be covered by each measurement is chosen according to the probability distribution weighted by the importance of each group, i.e., the larger the norm of the coefficients of the observables in the group is, the larger the probability becomes.  
As shown in previous literature \cite{Hadfield2022,Huang2021,Hillmich2021}, the LDF grouping successfully decreases the number of measurements in the VQE problems without increasing the depth of the circuit. 
 Due to the circuit synthesis in Section~\ref{section:circuit-synthesis}, $M_{k\ell}$ and $V_k$ are written as the sum of Pauli observables, and the LDF grouping is also usable for the evaluation of $V_k$ and $M_{k\ell}$. 
Thus, it can be a good benchmark for evaluating the shadow-based strategy. For a more detailed explanation of the LDF grouping, please see \cite{Hadfield2022}.

For the estimation of $M_{k\ell}$, there is no room for measurement optimization since $\nbb = 1$. To emphasize the difference in the performances of the above four estimation strategies of $V_k$, we 
assume that $M_{k\ell}$ can be exactly computed (which corresponds to the case when the number of measurements for estimating $\{M_{{k\ell}}\}$ is negligible).
We can obtain the exact value of $M_{k\ell}$ in Eq. \eqref{eq:mkl-real} by directly calculating $\bra{v_{\rm ref}}R_{k,i}^{\dagger}P_q R_{\ell,i^{\prime}} \ket{v_{\rm ref}}$ with the state vector representation.
Finally, throughout our numerical experiment, for simulating the quantum process, we use Qulacs \cite{Suzuki2021}.   
\subsection{Evaluation of the infidelity}
\label{section:exp-trace-distance}
Under the setting we described in Section~\ref{section:exp-settings}, we execute the RTE and ITE for the initial five steps such as $T=x \delta t$ $(x\in \{0,1,2,3,4,5\})$ where we set $\delta t = 0.01$,
and see how $\di$ evolves depending on the strategies to estimate $V_k$. 
For $T=0$, we have $\di = 0$. For computing $\ket{v(\vecthetaidealT)}$, we exactly compute $V$ by using the state vector representation, while for estimating $\ket{v(\vecthetaT)}$, we evaluate $V$ by using each measurement strategy. Given the total number of measurements for estimating each $V_k$ as $N_{\rm total}$, it is set to be $N_{\rm total} = 5\times \nba$  for each strategy
(thus, $N_{\rm shot} = 5\times \nba$ for the shadow-based strategy and $N_{\rm naive} =5$ for the naive strategy). 
For example, in the case of the ${\rm H}_2$ molecule Hamiltonian, $\nba = 184$, and the total number of measurements is $920$. 

\figfour{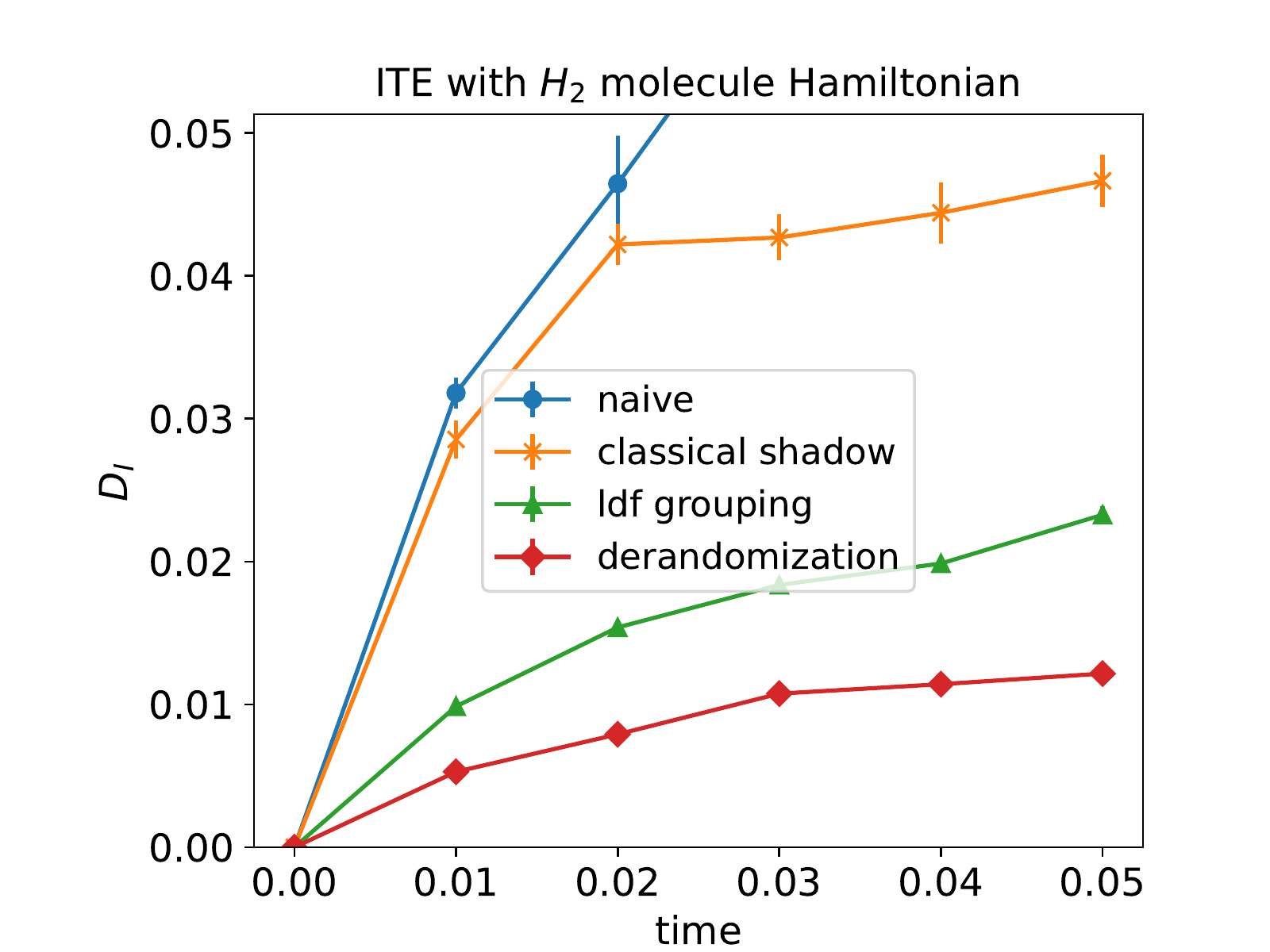}{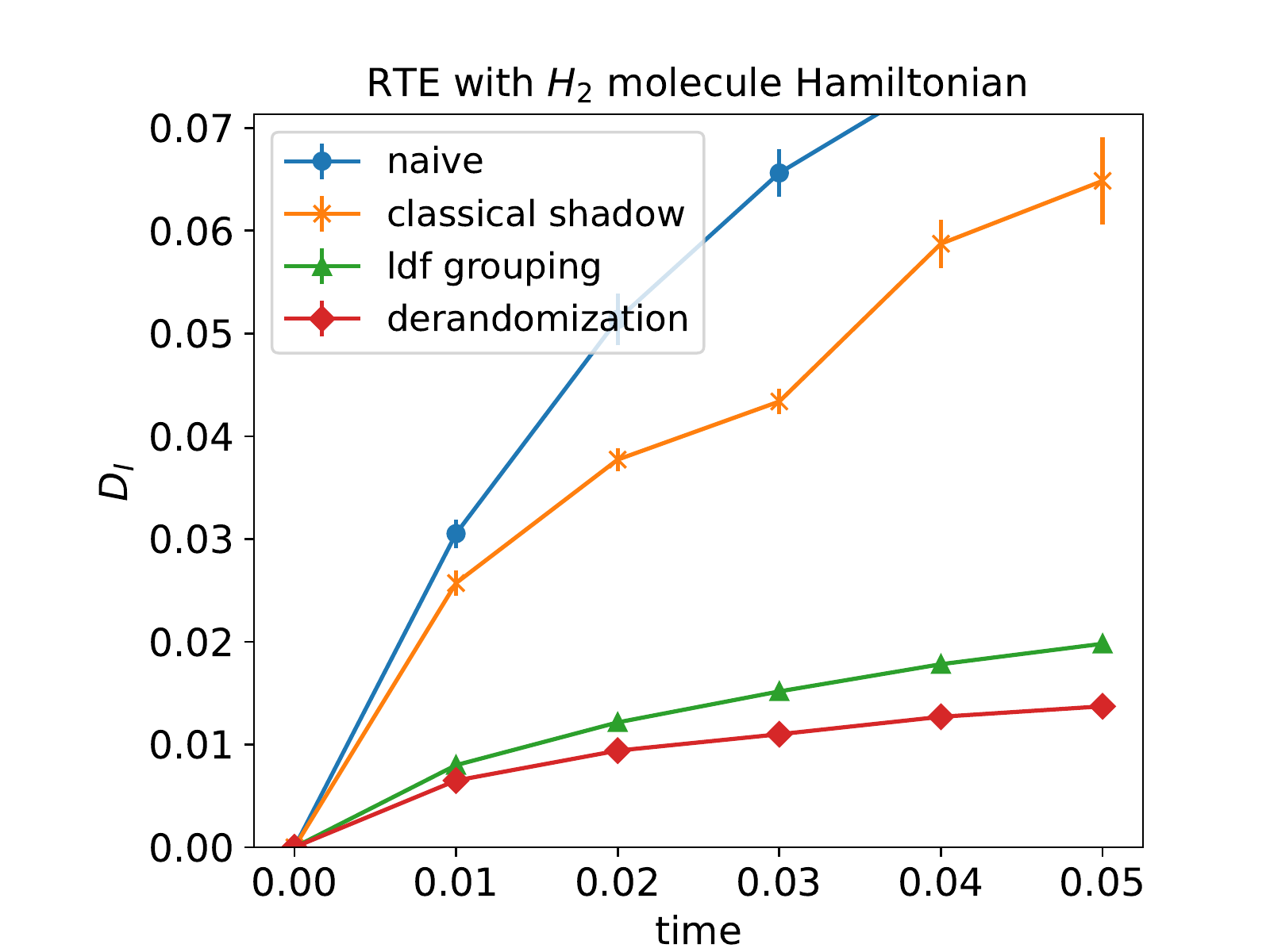}{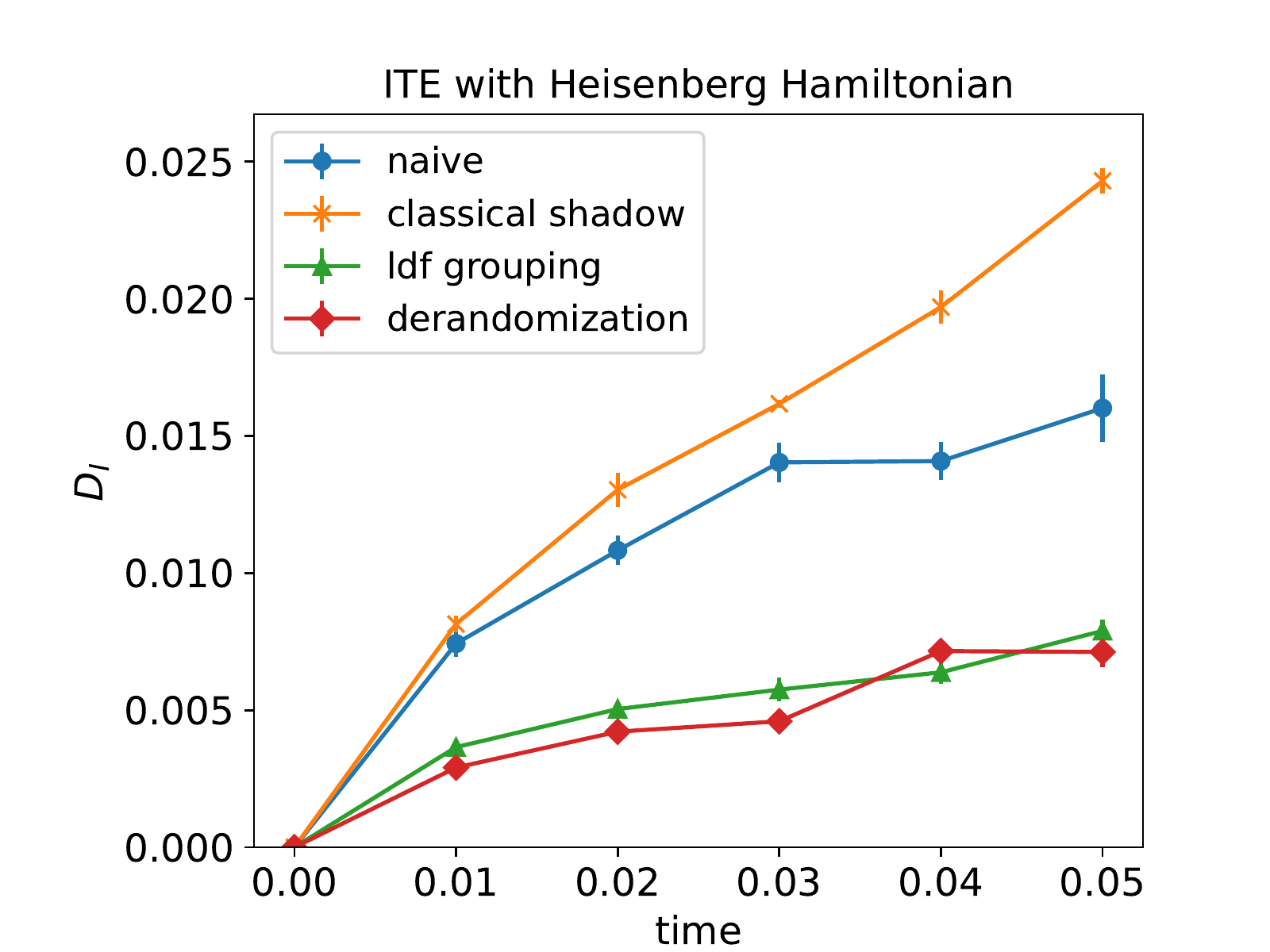}{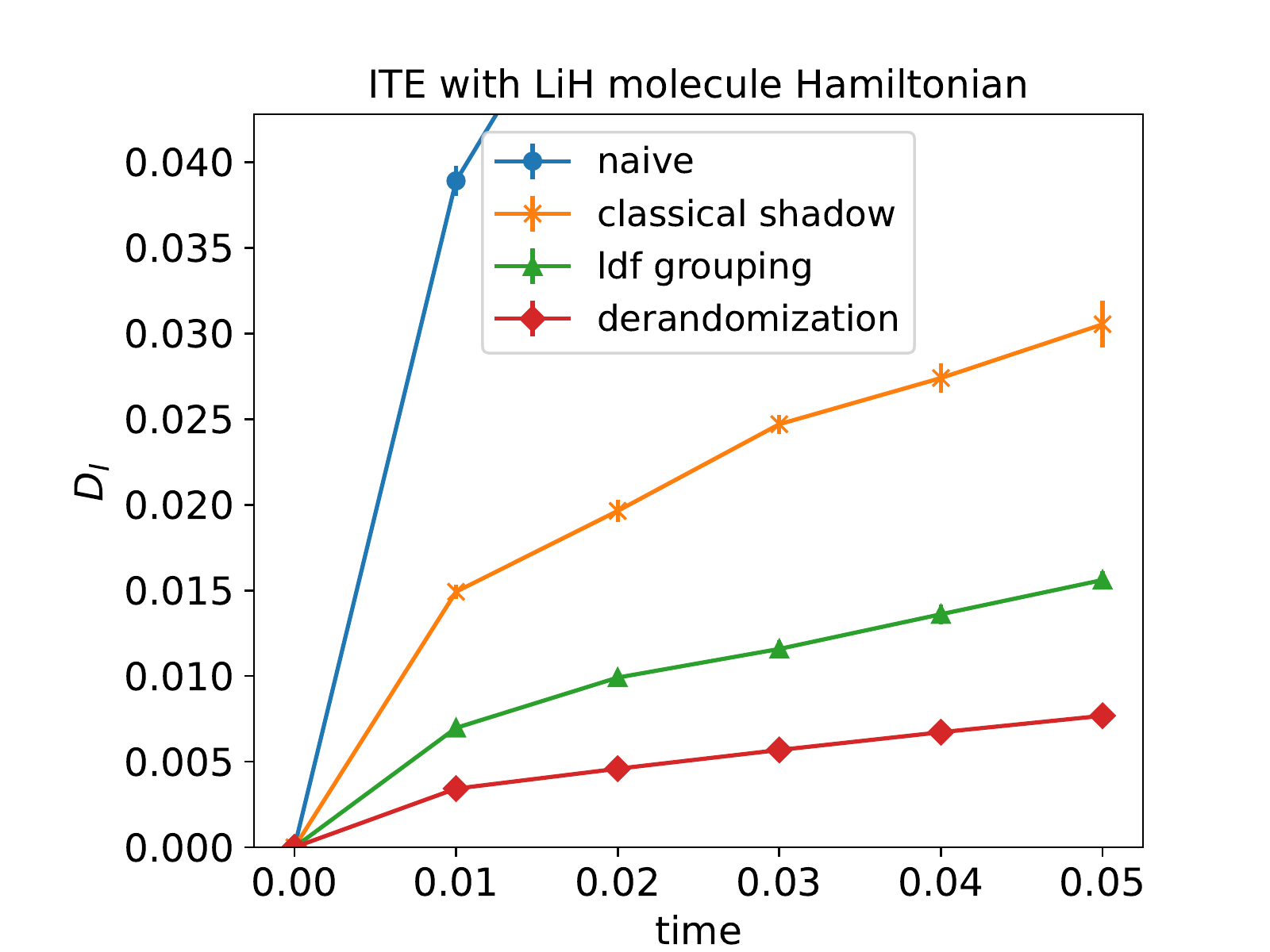}{220pt}{220pt}{The evolutions of $\di$ for each $T$. Four subfigures correspond to the followings: (a) ITE with ${\rm H}_2$ molecule Hamiltonian, (b) RTE with ${\rm H}_2$ molecule Hamiltonian, (c) ITE with Heisenberg model Hamiltonian, and (d) ITE with LiH molecule Hamiltonian.}{fig:time-plot}

The resulting evolutions of $\di$ for each $T$ are shown in Fig.~\ref{fig:time-plot}. There are four subfigures in it, which correspond to the followings: (a) ITE with ${\rm H}_2$ molecule Hamiltonian, (b) RTE with ${\rm H}_2$ molecule Hamiltonian, (c) ITE with Heisenberg model Hamiltonian and (d) ITE with LiH molecule Hamiltonian. We perform five trials and write the mean and std error of the mean for each data point. For each trial, different gate types are randomly chosen in PQC at initialization, but the chosen gate types are the same in every strategy as far as they are in the same trial. For example, the gate types chosen in the first trial of the naive strategy are the same as those in the first trial of the classical shadow strategy. 

In Fig.~\ref{fig:time-plot}, we see that the classical shadow outperforms the naive method other than the problems with the Heisenberg model Hamiltonian. 
The reason why the classical shadow does not work well in the case of the Heisenberg model Hamiltonian is as follows. The Heisenberg model Hamiltonian \eqref{eq:heisenberg-Hamiltonian} has $\nba = 24$ terms, and in the classical shadow, it holds
\begin{equation}
\label{eq:heisenberg-Hamiltonian-covering}
\left\langle\frac{1}{q(\tilde{P}_r)}\right\rangle_{\gv} \simeq  3^3,
\end{equation} 
where we use that the locality of most of the terms in \eqref{eq:heisenberg-Hamiltonian} is two and in the observable to evaluate $V_k$, the locality increases by one (see \eqref{eq:locality-increase}). 
By substituting these into \eqref{eq:condition-for-shadow-advantage}, 
\begin{equation}
		\frac{N_{\rm hybrid}^{\rm total}}{\numnaive^{\rm total}} 
	\lesssim \frac{1}{\nba } \times \left\langle\frac{1}{q(\tilde{P}_r)}\right\rangle_{\gv} \times \frac{9}{4}
	\simeq 2.5,
\end{equation} 
meaning that there is not an advantage of using the classical shadow algorithm in the Heisenberg model Hamiltonian as far as we see the upper bound. 

Even though the classical shadow algorithm successfully works in the other problems compared to the naive strategy, the LDF grouping outperforms the classical shadow in every problem. That is because the classical shadow contains inefficiency regarding the choice of the measurement basis. The most significant inefficiency might exist in the choice of the measurement basis for the ancilla qubit. For evaluating $V_k$, we need to evaluate the following summation of the observables: 
\begin{equation}
\label{eq:prs}
	\sum_r^{\nba} \alpha_r X \otimes P_r.
\end{equation}
Since all the observables have $X$ in the first qubit, the measurement basis that does not have $X$ in the first qubit does not give any information about $V_k$. Conversely, only the measurement basis that has $X$ in the first qubit covers the terms in \eqref{eq:prs}. However, the measurement basis is chosen with the probability of $1/3$, which is obviously inefficient in the computation of $V_k$. The inefficiency is the reason why the LDF grouping outperforms the classical shadow. 

We enjoy the benefit of the shadow-based strategy with a more intelligent selection of the measurement basis.
The derandomization, which optimizes the measurement basis so that $1/q(P_r)$ becomes smaller, outperforms the LDF grouping other than the problem with the Heisenberg model Hamiltonian. 
In the Heisenberg model Hamiltonian, all observables are covered by the only three measurement basis:
\begin{equation}
XXXXXXX, XYYYYYY, XZZZZZZ.
\end{equation}
Finding such a measurement basis is easy both for the LDF grouping and the derandomization. 
In our numerical
experiment, the LDF grouping successfully groups all the observables into three groups, where each group is covered by one of the above measurement basis. 
Thus, the measurements in the LDF grouping are nearly optimal and equal to the derandomization in the Heisenberg model Hamiltonian. However, in the other problems, the observables are not covered by a few measurement basis. Thus the derandomization, which is a relatively intelligent way of finding the measurement basis, outperforms the LDF grouping. 

We also compute $\di$ at $T=1$ with various total numbers of measurements $N_{\rm total}$. Similarly to Fig~\ref{fig:time-plot}, each data point is the mean and the standard error of the mean in five trials. The results are shown in Fig.~\ref{fig:shot-plot} (the labels of the subfigures are the same as those of Fig.~\ref{fig:time-plot}). We note that both the horizontal and vertical axes are in the logarithmic scale in the figure. We see that the same conclusion holds for the performance of each measurement strategy. Additionally, we see that $\di$ almost decreases with the square root of $N_{\rm total}$, which is consistent with \eqref{eq:bound-hybrid} in Appendix~\ref{section:shot-noise}.

\figfour{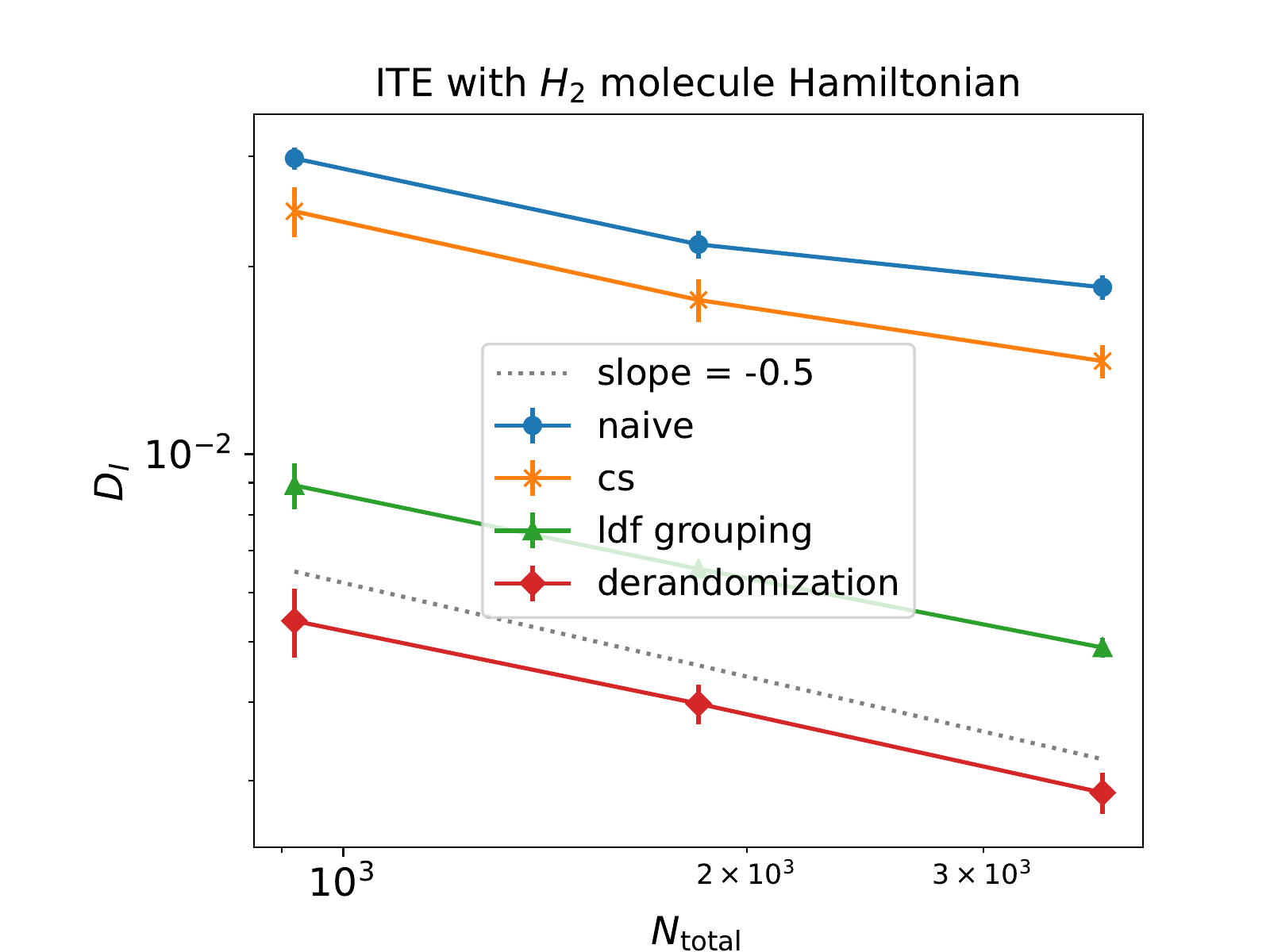}{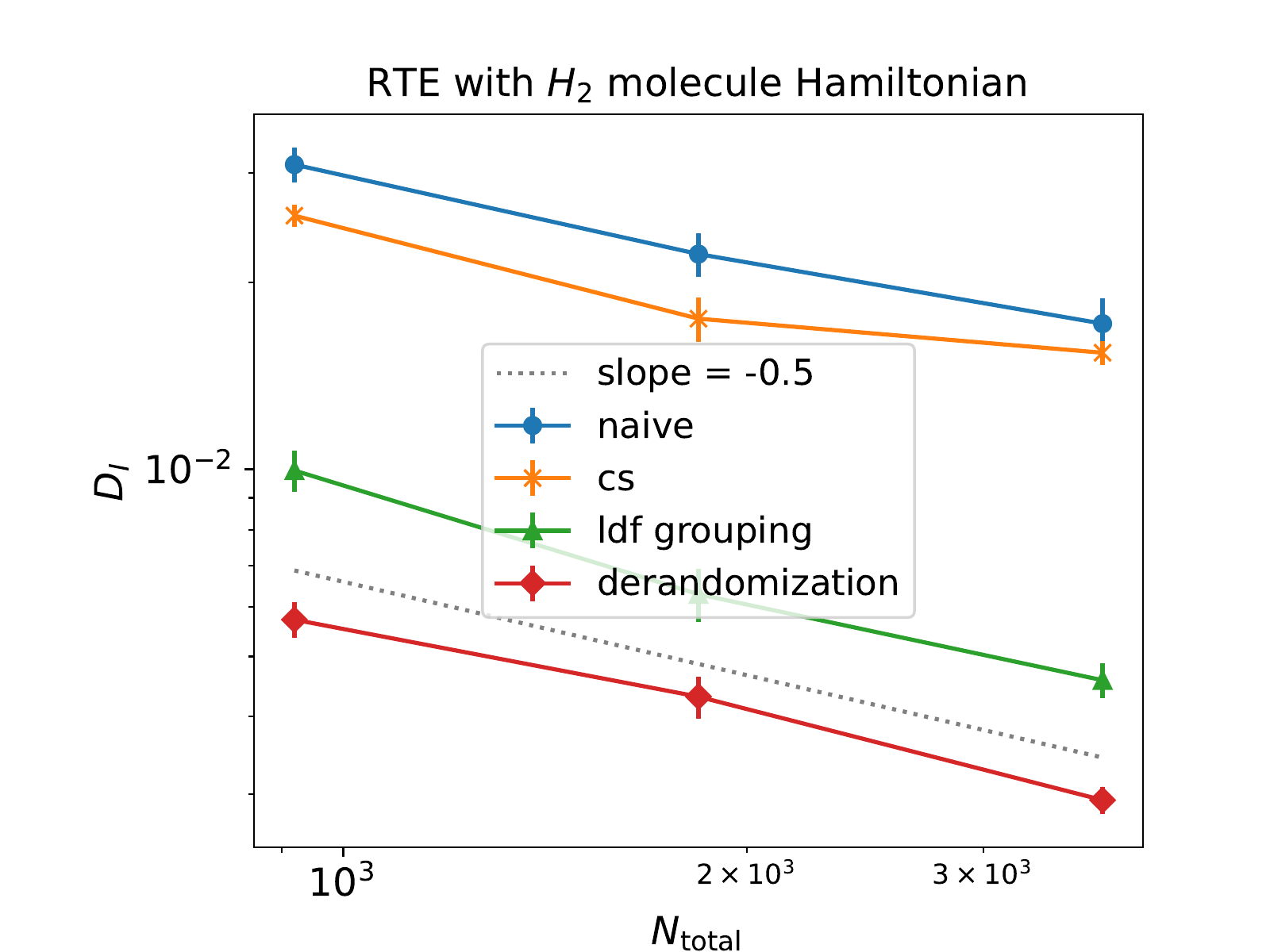}{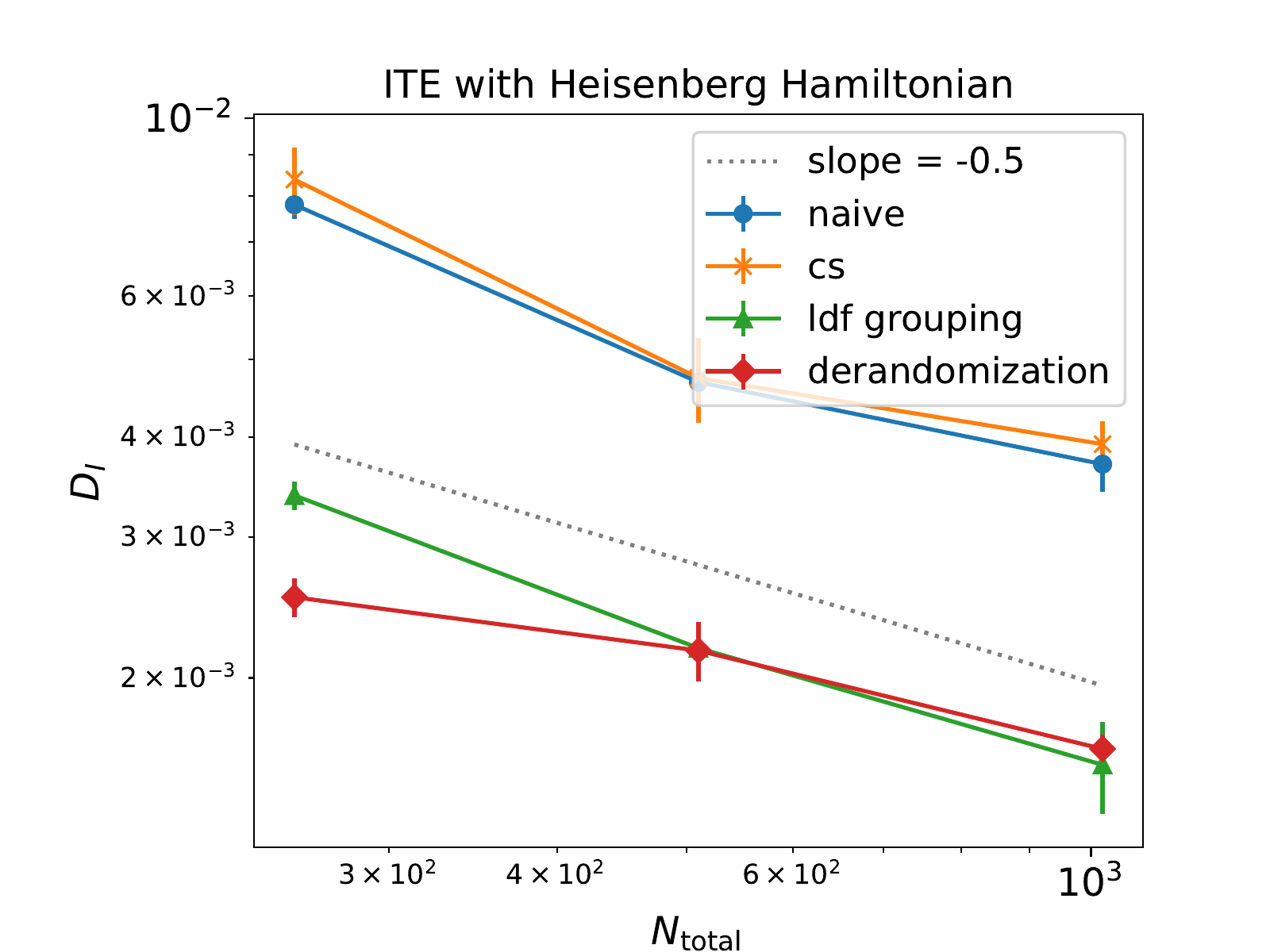}{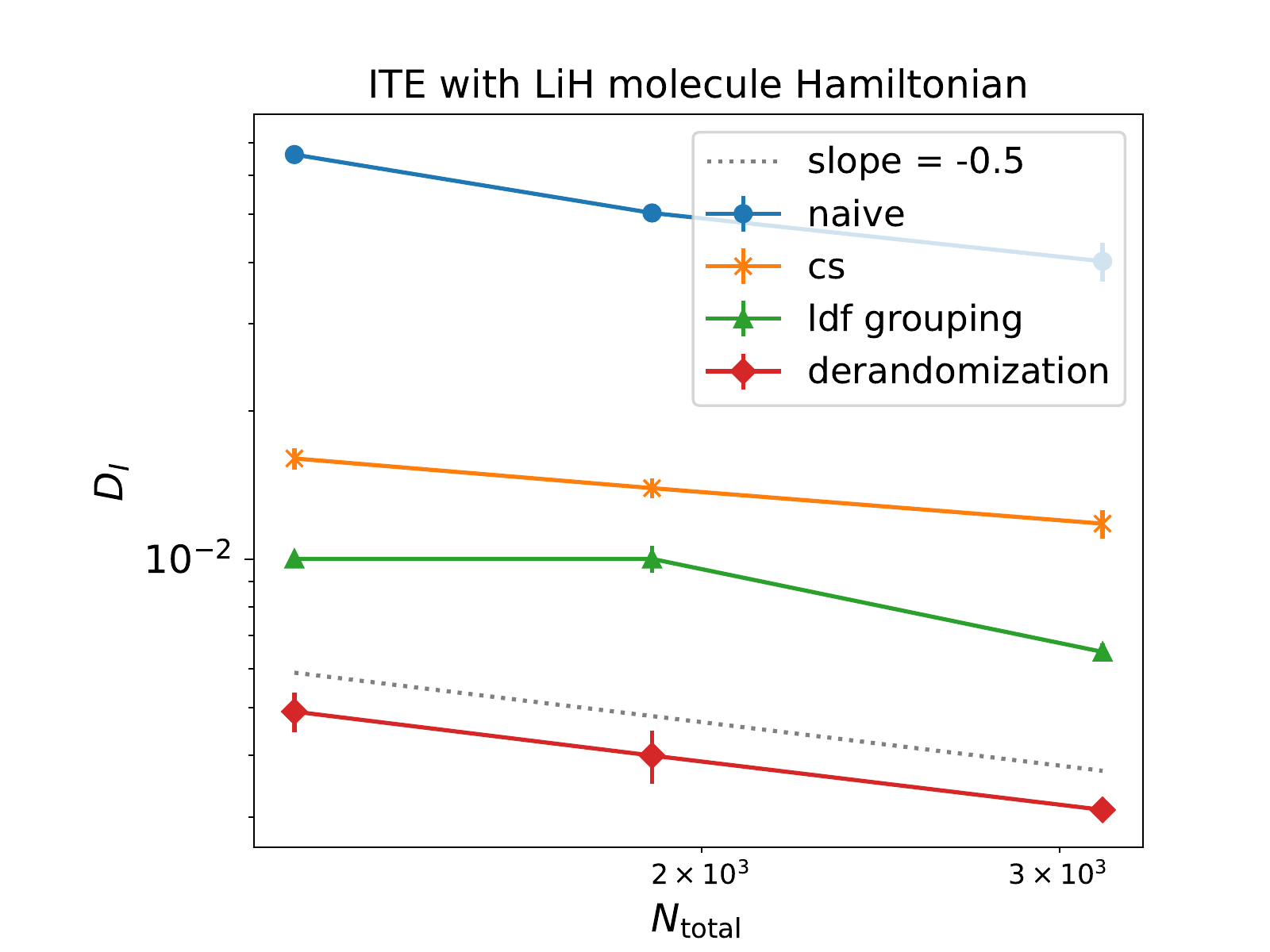}{220pt}{220pt}{The value of $\di$ at $T=1$ for each $N_{\rm total}$. As in Fig.~\ref{fig:time-plot}, four subfigures correspond to the followings: (a) ITE with ${\rm H}_2$ molecule Hamiltonian, (b) RTE with ${\rm H}_2$ molecule Hamiltonian, (c) ITE with Heisenberg model Hamiltonian, and (d) ITE with LiH molecule Hamiltonian. {
Both the horizontal and the vertical axes are in the logarithmic scale. 
		As a reference, we plot a slope proportional to $N_{\rm total}^{-0.5}$ by a gray dotted line. 
	}
}{fig:shot-plot}

\subsection{Evaluation of the variance}
\label{section:exp-variance}

Finally, we numerically compare the variance ${\rm Var}(\tilde{V}_k)$ and ${\rm Var}(\tilde{V}_k^{\rm naive})$ for a given total number of measurements. We set the total number of measurements as $5 \times \nba$.  We compute the variance in the two problems: ITE with ${\rm H}_2$ molecule Hamiltonian, which is shown in TABLE~\ref{table:variance-h}, and ITE with Heisenberg model Hamiltonian, which is shown in TABLE~\ref{table:variance-heisenberg}.

The detail of the contents of the tables is as follows. 
In the `Variance' column, we show the average of the variance; namely, 
\begin{equation}
{\rm Variance} = \frac{1}{N_P}\sum_{k=1}^{\numparams} {\rm Var}(\tilde{V}_k),
\end{equation}
for the shadow-based strategy and 
\begin{equation}
{\rm Variance} = \frac{1}{N_P}\sum_{k=1}^{\numparams} {\rm Var}(\tilde{V}_k^{\rm naive}),
\end{equation}
for the naive strategy. 
Additionally, checking
\begin{equation}
		(\Delta V_k)^2 \lesssim \sum_{i=1}^{\numderiv}  \sum_{r=1}^{\nba}  \frac{(\gv)^2}{\qfunc{P_r}},~\left(\Delta V_k^{\rm naive}\right)^2 \simeq \sum_{i=1}^{\numderiv} \sum_{r=1}^{\nba}  (\gv)^2, 
\nonumber
\end{equation}
in the `Approximation' column, 
we show the approximate value of `Variance', which is denoted by 
\begin{equation}
	{\rm Approximation} =\frac{1}{N_P}\sum_{k=1}^{\numparams}
 \left(  \frac{1}{N_{\rm shot}} \sum_{i=1}^{\numderiv} \sum_{r=1}^{\nba}  \frac{(\gv)^2}{\qfunc{P_r}}\right)  ,
\end{equation}
for the shadow-based strategy, and 
\begin{equation}
	{\rm Approximation} = \frac{1}{N_P} \sum_{k=1}^{\numparams}\left( \frac{1}{N_{\rm naive}}\sum_{i=1}^{\numderiv} 
  \sum_{r=1}^{\nba} (G_{k,i,r})^2\right),
\end{equation} 
for the naive strategy. Finally, in the `Diff' row, we show the averege difference between the true value and the approximate value of the variance, namely, 
\begin{equation}
{\rm Diff} = \frac{1}{\numparams} \sum_{k=1}^{\numparams} \left|
{\rm Var}(\tilde{V}_k) - \frac{1}{\ns} \sum_{i=1}^{\numderiv} \sum_{r=1}^{\nba}  \frac{(\gv)^2}{\qfunc{P_r}} 
\right|,
\end{equation} 
for the shadow-based strategy, and 
\begin{equation}
{\rm Diff} = \frac{1}{\numparams} \sum_{k=1}^{\numparams} \left|
 {\rm Var}(\tilde{V}_k^{\rm naive}) -
 \frac{1}{N_{\rm naive}}
  \sum_{i=1}^{\numderiv} 
  \sum_{r=1}^{\nba} (G_{k,i,r})^2 \right|,
\end{equation} 
for the naive strategy. 
The value of `Variance' depends on the parameters of the ansatz. We choose parameters in two different ways when computing `Variance': (i) all zero, where all parameters are set to be zero, and (ii) random, where all parameters are randomly selected. For the random parameters choice case, we generate five patterns of parameters. 

There are several implications from the results in the tables. 
First, both in TABLE \ref{table:variance-h} and TABLE \ref{table:variance-heisenberg}, we see that all `Variance' are almost equal to the corresponding `Approximation', which shows the validity of the approximation in \eqref{eq:delv-bound} and \eqref{eq:delv-naive-bound} in Section~\ref{section:algorithm-advantage}. 
Second, the variances are consistent with the results in Section~\ref{section:exp-trace-distance}. Namely, 
the smaller variance results in the smaller infidelity in Fig.~\ref{fig:time-plot} and Fig.~\ref{fig:shot-plot}. Finally, variances in the derandomization are much smaller than those in the other measurement methods. Particularly, the large difference in the variances between the classical shadow and the derandomization shows the importance of optimizing the covering probability $q(P_r)$ for decreasing the variance (hence the infidelity) and implies that other shadow-based strategies \cite{Hadfield2022,Hillmich2021}, which also optimizes the covering probability, also achieve good performance. Studying the effect of using \cite{Hadfield2022,Hillmich2021} in VQS is left for future work. 

\begin{table}[]
\caption{Variance, Approximation, and Diff in the problem of ITE with ${\rm H}_2$ molecule Hamiltonian.}
\label{table:variance-h}
\begin{tabular}{p{0.10\textwidth} p{0.30\textwidth} p{0.18\textwidth} p{0.18\textwidth} p{0.18\textwidth}}
\hline\hline
Parameters & Measurement & Variance & Approximation  & Diff \\
\hline
\multirow{3}{*}{all zero} &\textbf{naive}   & 0.731    &   0.743    &   0.012              \\

&\textbf{classical shadow} &   0.524    &  0.525     &    0.001     \\

&\textbf{derandomization}  &  0.020     &  0.020      &   0.000           \\ 
\hline
 & \textbf{naive} &0.741 & 0.743 & 0.001 \\
random & \textbf{classical shadow} &0.525 & 0.525 & 0.000 \\
(pattern1)& \textbf{derandomization} &0.020 & 0.020 & 0.000 \\
\hline
& \textbf{naive} &0.731 & 0.743 & 0.011 \\
random& \textbf{classical shadow} &0.528 & 0.525 & 0.002 \\
(pattern2)& \textbf{derandomization} &0.019 & 0.020 & 0.000 \\
\hline
& \textbf{naive} &0.733 & 0.743 & 0.010 \\
random & \textbf{classical shadow} &0.526 & 0.525 & 0.000 \\
(pattern3)& \textbf{derandomization} &0.020 & 0.020 & 0.000 \\
\hline
& \textbf{naive} &0.736 & 0.743 & 0.006 \\
random & \textbf{classical shadow} &0.528 & 0.525 & 0.002 \\
(pattern4)& \textbf{derandomization} &0.020 & 0.020 & 0.000 \\
\hline
& \textbf{naive} &0.732 & 0.743 & 0.010 \\
random & \textbf{classical shadow} &0.510 & 0.525 & 0.014 \\
(pattern5) & \textbf{derandomization} &0.019 & 0.020 & 0.000 \\
\hline
\end{tabular}
\end{table}

\begin{table}[]
\caption{Variance, Approximation, and Diff in the problem of ITE with Heisenberg model Hamiltonian.}\label{table:variance-heisenberg}
\begin{tabular}{p{0.10\textwidth} p{0.30\textwidth} p{0.18\textwidth} p{0.18\textwidth} p{0.18\textwidth}}
\hline\hline
Parameters & Measurement & Variance & Approximation  & Diff \\
\hline
& \textbf{naive} &0.0475 & 0.0480 & 0.0004 \\
all zero & \textbf{classical shadow} &0.0457 & 0.0450 & 0.0007 \\
& \textbf{derandomization} &0.0063 & 0.0059 & 0.0004 \\
\hline
& \textbf{naive} &0.0469 & 0.0480 & 0.0010 \\
random & \textbf{classical shadow} &0.0464 & 0.0450 & 0.0014 \\
(pattern1) & \textbf{derandomization} &0.0055 & 0.0059 & 0.0003 \\
\hline
& \textbf{naive} &0.0472 & 0.0480 & 0.0007 \\
random & \textbf{classical shadow} &0.0436 & 0.0450  & 0.0013 \\
(pattern2) & \textbf{derandomization} &0.0054 & 0.0059 & 0.0004 \\
\hline
& \textbf{naive} &0.0471 & 0.0480 & 0.0008 \\
random & \textbf{classical shadow} &0.0443 & 0.0450 & 0.0006 \\
(pattern3)& \textbf{derandomization} &0.0054 & 0.0059 & 0.0005 \\
\hline
& \textbf{naive} &0.0467 & 0.0480 & 0.0012 \\
random & \textbf{classical shadow} &0.0449 & 0.0450 & 0.0000 \\
(pattern4) & \textbf{derandomization} &0.0065 & 0.0059 & 0.0006 \\
\hline
& \textbf{naive} &0.0469 & 0.0480 & 0.0010 \\
random & \textbf{classical shadow} &0.0456 & 0.0450 & 0.0006 \\
(pattern5) & \textbf{derandomization} &0.0058 & 0.0059 & 0.0001 \\
\hline
\end{tabular}
\end{table}

\section{Discussion}
\label{section:conclusion}
The variational quantum simulation (VQS) successfully gives a way of simulating quantum systems in those devices, which is considered as one of the practical applications the near-term quantum devices. However, when applying VQS to a large quantum system where we have practical interests, the number of running a quantum device, i.e., the number of measurements, dramatically increases to reduce the effect of shot noise. Even though the grouping methods and the shadow-based strategies have recently been proposed to decrease the number of measurements in the variational quantum optimization (VQO), they are not utilized in VQS because the way of evaluating the observables in VQS is different from the one in VQO and we cannot straightforwardly apply the shadow-based strategies to VQS.  
 In this paper, by changing the circuit and observables
 in VQS, we make it possible to apply shadow-based strategies and grouping methods to VQS. We particularly focus on the shadow-based strategies; we theoretically show that with the shadow-based strategy, we can
greatly decrease the number of measurements in VQS compared to the naive strategy, which has been utilized since the original proposal of VQS \cite{Li2017,mcardle2019variational,yuan2019theory}. 
Our numerical demonstration of ITE and RTE using various Hamiltonians, including chemical Hamiltonians, also implies the validity of using the shadow-based strategy. Particularly, the derandomization \cite{Huang2021}, one of the shadow-based strategies, achieves the best performance in most of the Hamiltonians.

Though we examine only the classical shadow and the derandomization in our numerical experiments as the shadow-based strategies, other shadow-based strategies \cite{Hillmich2021,Wu2021} can also be utilized to reduce the number of measurements in VQS. For example, the literature \cite{Hillmich2021} shows that their proposed shadow-based strategy using decision diagrams effectively reduces the estimator's variance to smaller than that in the LDF grouping, which achieves the second-best performance in our numerical demonstration. 
Additionally, as we note in Section~\ref{section:introduction}, recent literature \cite{yen2022deterministic}
reports that their grouping strategies outperform shadow-based strategies, such as derandomization, by optimizing parameters in the grouping so that the variance of the observable is minimized. 
Studying the effect of using those shadow-based strategies and the grouping strategies in VQS is a good direction for future work. We believe that our work takes the first step toward measurement optimization in VQS and paves the way for building practical applications for simulating quantum systems. 

\section*{Acknowledgement}
This work was supported by Grant-in-Aid for JSPS Research Fellow 22J01501 and by Leading Initiative for Excellent Young Researchers MEXT Japan and JST presto
(Grant No. JPMJPR1919) Japan. 
This work was supported by MEXT Quantum Leap Flagship Program (MEXT Q-LEAP) Grant Number JPMXS0120319794.
This paper is partly
based on results obtained from a project, JPNP16007,
commissioned by the New Energy and Industrial Technology Development Organization (NEDO), Japan.

\appendix
\section{Statistical properties of $\nu$}
\label{section:statistics}

\subsubsection*{Proof that $\nu$ is an unbiased estimator of ${\rm Tr}(H\rho)$}
In this Section, we show that $\nu$ is the unbiased estimator of ${\rm Tr}(H\rho)$. When measurements are chosen probabilistically, 
\begin{equation}
\begin{split}	
\expect{\nu_r} &= \sum_{j=1}^K a_j \frac{1}{\qfunc{P_j}} \sum_{M_r \in \{X,Y,Z\}^{\otimes n}} f(P_j, M_r) q_{\rm{m}}(M_r) \sum_{b} p(\rho, M_r; b)\mu(P_j, b) \\
    &= \sum_{j=1}^K a_j \frac{1}{\qfunc{P_j}} \sum_{M_r \in \{X,Y,Z\}^{\otimes n}} f(P_j, M_r) q_{\rm{m}}(M_r) {\rm Tr}(P_j \rho) \\
    &= \sum_{j=1}^K a_j {\rm Tr}(P_j\rho)  \\
    &= {\rm Tr}(\mathcal{H} \rho),
\end{split}
\end{equation}
where we use \eqref{eq:cover-exp} in the second equality, and we use \eqref{eq:covering-probability-probabilistic} in the third equality. Thus, $\expect{\nu} = \sum_{r=1}^{\ns} \expect{\nu_r}/N_{\rm shot} = {\rm Tr}(\mathcal{H\rho})$.

When  measurements are chosen deterministically, 
\begin{equation}
\begin{split}	
	\expect{\nu_r} &= \sum_{j=1}^K a_j \frac{1}{\qfunc{P_j}} f(P_j, M_r) \sum_b p(\rho, M_r; b)\mu(P_j, b) \\
	&= \sum_{j=1}^K a_j \frac{1}{\qfunc{P_j}} f(P_j, M_r) {\rm Tr}(P_j\rho)\\
	&= \sum_{j=1}^K a_j \frac{\ns f(P_j, M_r){\rm Tr}(P_j\rho)}{\sum_{r^{\prime}=1}^{\ns} f(P_j, M_{r^{\prime}})}, 
\end{split}
\end{equation}
where we use \eqref{eq:covering-probability-deterministic} in the last equality.
Therefore,  
\begin{equation}
\begin{split}	
	\expect{\nu} &= \frac{1}{\ns}\sum_{r=1}^{\ns} \expect{\nu_r} \\
	&= \sum_{j=1}^K a_j \frac{\sum_{r=1}^{\ns} f(P_j, M_r){\rm Tr}(P_j\rho)}{\sum_{r^{\prime}=1}^{\ns} f(P_j, M_{r^{\prime}})} \\
	&= \sum_{j=1} a_j{\rm Tr}(P_j\rho) \\
	&= {\rm Tr}(\mathcal{H}\rho).
\end{split}
\end{equation}

\subsubsection*{Variance of $\nu$}
The variance of $\nu$ is defined by
\begin{align}
\label{eq:variance-def}
	\var(\nu) = \expect{\nu^2} - (\expect{\nu})^2 
	= \frac{1}{\ns}\left( \frac{1}{\ns}\sum_{r=1}^{\ns} \expect{\nu_r^2} - [{\rm Tr}(\mathcal{H}\rho)]^2 \right).
\end{align}
When measurements are probabilistically chosen, 
\begin{align}	
\label{eq:nu-square}
	\expect{\nu_r^2} = \sum_{j=1}^K\sum_{\ell=1}^K a_j a_{\ell}
	\sum_{M_r \in \{X, Y, Z\}^{\otimes n}} \frac{f(P_j, M_r)}{\qfunc{P_j}}  \frac{f(P_{\ell}, M_r)}{q(P_{\ell})} q_m(M_r) \sum_{b_r\in \{-1,1\}^{\otimes n}} p(\rho, M_r; b_r)\mu(P_j, b_r)\mu(P_{\ell}, b_r).
\end{align}
As long as $M_r$ covers both $P_j$ and $P_{\ell}$
\begin{equation}
\label{eq:pjpl-rho}
 \sum_{b_r\in \{-1,1\}^{\otimes n}} p(\rho, M_r; b_r)\mu(P_j, b_r)\mu(P_{\ell}, b_r) = {\rm Tr}(P_j P_{\ell} \rho).
\end{equation}
Substituting \eqref{eq:pjpl-rho} to \eqref{eq:nu-square}, we obtain 
\begin{equation}
	\expect{\nu_r^2} = \sum_{j=1}^K\sum_{\ell=1}^K a_j a_{\ell} 
	 \sum_{M_r \in \{X, Y, Z\}^{\otimes n}} \frac{f(P_j, M_r)}{\qfunc{P_j}}  \frac{f(P_{\ell}, M_r)}{q(P_{\ell})} q_m(M_r)
	{\rm Tr}(P_j P_{\ell}\rho).
\end{equation}
By using $f(P_j, M_r)^2 = f(P_j, M_r) $, 
\begin{equation}
\label{eq:diagonal-term}
	\sum_{M_r \in \{X, Y, Z\}^{\otimes n}} \left(\frac{f(P_j, M_r)}{\qfunc{P_j}}\right)^2  q_m(M_r)  =\frac{1}{\qfunc{P_j}},
\end{equation}
and therefore, 
{\begin{equation}
	\expect{\nu_r^2} = \sum_{j=1}^K \frac{a_j^2}{\qfunc{P_j}} + \sum_{j\neq \ell} a_j a_{\ell} \sum_{M_r \in \{X, Y, Z\}^{\otimes n}} \frac{f(P_j, M_r)}{\qfunc{P_j}}  \frac{f(P_{\ell}, M_r)}{q(P_{\ell})}q_m(M_r){\rm Tr}(P_j P_{\ell}\rho).
\end{equation}
As a result, 
\begin{equation}
	\var(\nu) = \frac{1}{\ns}\left[
		 \sum_{j=1}^K \frac{a_j^2}{\qfunc{P_j}} + \sum_{j\neq k} a_j a_{\ell} 
		 \sum_{M_r \in \{X, Y, Z\}^{\otimes n}}\frac{f(P_j, M_r)}{\qfunc{P_j}}  \frac{f(P_{\ell}, M_r)}{q(P_{\ell})} q_m(M_r)
		 {\rm Tr}(P_j P_{\ell}\rho) - [{\rm Tr}(\mathcal{H}\rho)]^2
	\right].
\end{equation}

	In the deterministic case, by using \eqref{eq:pjpl-rho}, 
\begin{align}
	\frac{1}{\ns}\sum_{r=1}^{\ns}\expect{\nu_r^2} = \frac{1}{\ns}\sum_{j=1}^K \sum_{\ell=1}^K a_j a_{\ell} \sum_{r=1}^{\ns} \frac{f(P_j, M_r)}{\qfunc{P_j}}  \frac{f(P_{\ell}, M_r)}{\qfunc{P_{\ell}}} {\rm Tr}(P_j P_{\ell}\rho). 
\end{align}	
By using a calculation similar to
\eqref{eq:diagonal-term}, we can separate the diagonal term and obtain
\begin{equation}
 	\var(\nu) = \frac{1}{\ns} \left[
 		 \sum_{j=1}^K \frac{a_j^2 }{q(P_j)} +\frac{1}{\ns} \sum_{j\neq \ell} a_j a_{\ell} \sum_{r=1}^{\ns}\frac{f(P_j, M_r)}{\qfunc{P_j}}  \frac{f(P_{\ell}, M_r)}{\qfunc{P_{\ell}}} {\rm Tr}(P_j P_{\ell} \rho) - [{\rm Tr}(\mathcal{H}\rho)]^2
 	\right].
 \end{equation}

\section{Variance in classical shadow}
\label{section:variance-cs}
Here we derive the variance of $\nu$ in the classical shadow \cite{Huang2020}.
From \eqref{eq:g-probabilistic}, we obtain
\begin{align}
\label{eq:cs-gpjpl}
	g(P_j, P_\ell) &= 3^{{\rm locality}(P_j)} 	3^{{\rm locality}(P_{\ell})} 
\left(\frac{1}{3^n}	 \sum_{M_r \in \{X, Y, Z\}^{\otimes n}} 
	f(P_j, M_r) 	f(P_{\ell}, M_r)\right) \\
	&= 3^{{\rm locality}(P_j)} 	3^{{\rm locality}(P_{\ell})} 
	 \prod_{i=1}^n 
	 \frac{1}{3} \sum_{M_r[i]\in \{X, Y, Z\}} f_{\rm local}(P_j[i], M_r[i]) f_{\rm local}(P_{\ell}[i], M_r[i]).\label{eq:cs-gpjplymadd}
\end{align} 
We can compute the value inside the product as 
\begin{align}
 \frac{1}{3} \sum_{M_r[i]\in \{X, Y, Z\}} f_{\rm local}(P_j[i], M_r[i]) f_{\rm local}(P_{\ell}[i], M_r[i])
 = 
 \left\{
 \begin{array}{cl}
 \label{eq:cs-expand-flocal}
 	1 & P_j[i] = P_{\ell}[i] = I \\
 	1/3 & (P_j[i] = I~{\rm or}~P_{\ell}[i] = I)~{\rm and}~P_j[i]\neq P_{\ell}[i] \\
    1/3 & P_j[i] = P_{\ell}[i] \neq I \\
    0  & otherwise
 \end{array}. 
 \right.
\end{align}
Also, we can expand
\begin{equation}
\label{eq:cs-expand-locality}
	3^{{\rm locality}(P_j)} 	3^{{\rm locality}(P_{\ell})}
	= \prod_{i=1}^n c_{\rm local}(P_j[i], P_{\ell}[i]), 
\end{equation}
where 
\begin{equation}
	c_{\rm local}(P_j[i], P_{\ell}[i]) = 
	\left\{\begin{array}{cl}
		1 & P_j[i] = P_{\ell}[i] = I \\
		3 & (P_j[i] = I~{\rm or}~P_{\ell}[i] = I)~{\rm and}~P_j[i]\neq P_{\ell}[i] \\
		9 & P_j[i] = P_{\ell}[i] \neq I
	\end{array}\right. .
\end{equation}

By substituting \eqref{eq:cs-expand-flocal} and \eqref{eq:cs-expand-locality} into \eqref{eq:cs-gpjplymadd}, we obtain 
\begin{align}
	g(P_j, P_{\ell}) = g_{\rm cs}(P_j, P_{\ell})\equiv\prod_{i=1}^n g_{\rm local}(P_j[i], P_{\ell}[i]), 
\end{align}
where 
\begin{equation}
	g_{\rm local}(P_j[i], P_{\ell}[i]) = 
	\left\{
	\begin{array}{cl}
	1	& P_j[i] = I~{\rm or}~P_{\ell}[i] = I \\
	3 & P_j[i] = P_{\ell}[i] \neq I \\
	0 & otherwise
	\end{array}
	\right. .
\end{equation}
Consequently, 
\begin{equation}
	\var(\nu) = \frac{1}{\ns}\left[
		 \sum_{j=1}^K a_j^2 3^{{\rm locality}(P_j)}+ \sum_{j\neq \ell} a_j a_{\ell} g_{\rm cs}(P_j, P_{\ell}){\rm Tr}(P_j P_{\ell}\rho) - [{\rm Tr}(\mathcal{H}\rho)]^2
	\right].
\end{equation}


\section{Derivation of the bound for shot noise}
\label{section:shot-noise}
In \cite{Li2017}, the following infidelity 
\begin{equation}
\di = \sqrt{1 - 	\left|\left\langle v(\vecthetaidealT)\right| \left. v(\vecthetaT)\right\rangle\right|^2} \nonumber
\end{equation}
is evaluated when not using the shadow-based strategy. In this section, we firstly show the value of the previously computed $D_I$ and next, we evaluate $D_I$ when using the shadow-based strategy.

Suppose that there are $N$-steps for the update of the parameters, i.e., $N\delta t = T$, and let us write the state evolved by \eqref{eq:distored-time-evolution} and the ideal state evolved by \eqref{eq:ideal-time-evolution} at $r$-th step as $\ket{v^r}$ and $\ket{v_{\rm ideal}^r}$. Then, $\ket{v^N} = \ket{v(\vecthetaT)}$ and $\ket{v^N_{\rm ideal}} = \ket{v(\vecthetaidealT)}$. We also assume that $|v^0\rangle = |v^0_{\rm ideal}\rangle$.
By using the triangle inequality, we obtain 
\begin{equation}
\label{d-inequality}
\begin{split}	
	D_I\left(\ket{v^{r}_{\rm ideal}}, \ket{v^{r}}\right) &= 	D_I\left(U_r\ket{v^{r-1}_{\rm ideal}}, \ket{v^{r}}\right) \\
	&\leq D_I \left(U_r\ket{v_{\rm ideal}^{r-1}}, U_r \ket{v^{r-1}}\right) + D_I \left(U_r\ket{v^{r-1}}, \ket{v^r}\right) \\
	&= D_I \left(\ket{v_{\rm ideal}^{r-1}}, \ket{v^{r-1}}\right) + D_I \left(\ket{v_0^r}, \ket{v^r}\right), 
\end{split}
\end{equation}
for $r \geq 1$, 
where $U_r$ is the unitary transformation correspoinding to the $r$-th ideal update of the parameters by \eqref{eq:ideal-time-evolution}, and $\ket{v_0^r} \equiv U_r\ket{v^{r-1}}$. By applying \eqref{d-inequality} repeatedly, we obtain 
\begin{equation}
		D_I\left(\ket{v^{N}_{\rm ideal}}, \ket{v^{N}}\right) \leq \sum_{r=1}^N D_I \left(\ket{v_0^{r}}, \ket{v^r}\right). 
\end{equation}
Let 
\begin{equation}
	\delta {\vec{\dot{\theta}}^r}  \equiv 
	\left.\deriv{\vec{\theta}(t)}{t}\right|_{t=r\delta t} -
		\left.\deriv{\vec{\theta}_{\rm ideal}(t)}{t}\right|_{t=r\delta t}.
\end{equation}
Then, the direct calculation shows \cite{Li2017} 
\begin{equation}
	D_I (|v_0^r\rangle, |v^r\rangle) = \sqrt{\delta ({\vec{\dot{\theta^r}}})^T W^r \delta {\vec{\dot{\theta^r}}} (\delta t)^2 + O((\delta t)^3)}, 
\end{equation}
where each element of the matrix $W^r$ is given by
\begin{equation}
	W_{jk}^r = \deriv{\bra{v^{r-1}}}{\theta_j} \deriv{\ket{v^{r-1}}}{\theta_k} 
	- \deriv{\bra{v^{r-1}}}{\theta_j} \ket{v^{r-1}}\bra{v^{r-1}}\deriv{\ket{v^{r-1}}}{\theta_k}.
\end{equation}
Here, we use $\deriv{\langle v^{r-1}|}{t}|v^{r-1}\rangle =-\langle v^{r-1}|\deriv{|v^{r-1}\rangle }{t}$
, which is derived from $\deriv{}{t}\langle v^{r-1}|v^{r-1}\rangle
=0$.
By ignoring the $O((\delta t)^3)$ term inside the square root function, we obtain 
\begin{equation}
\begin{split}	
D_I\left(\ket{v^{N}_{\rm ideal}}, \ket{v^{N}}\right)
&\lesssim 
\sum_{r=1}^N \delta t\sqrt{
\delta ({\vec{\dot{\theta^r}}})^T W^r \delta {\vec{\dot{\theta^r}}}
} \\
&\le  T \sqrt{||W||_{\rm max}} ||\delta {\vec{\dot{\theta}}}||_{\rm max},
\end{split}
\end{equation}
where $||W||_{{\rm max}} = \max_{r}||W^r||_F$ and 
$||\delta \dot{\vec{\theta}}||_{\rm max} = \max_{0\leq t\leq T} ||\delta{\vec{\dot{\theta(t)}}}||_2$.

The value $\deldottheta$ is related with the amount of the shot noise. 
By substituting 
$(M + \delta M)^{-1} = M^{-1} - M^{-1}\delta M M^{-1} + O((\delta M)^2)$
in \eqref{eq:distored-time-evolution}, we obtain
\begin{equation}
	\deriv{\vec{\theta}(t)}{t} \simeq (M^{-1}\vec{V} + M^{-1}\delta\vec{V} - M^{-1}\delta M M^{-1} \vec{V}),  
\end{equation}
where we ignore the higher order terms of $\delta M$ and $\delta V$ with the assumption that $\delta M$ and $\delta V$ are sufficiently small. 
Then, 
\begin{equation}
\begin{split}	
	 |\deldottheta| &=  \left | 	\deriv{\vec{\theta}(t)}{t} - M^{-1}\vec{V} \right|\\
	 &\simeq  
	\left|M^{-1}\delta\vec{V} - M^{-1}\delta M M^{-1}\vec{V}\right| \\
	&\leq || M^{-1} ||_F ||\delta \vec{V}||_2 + ||M^{-1}||_F^2 ||\delta M||_F ||\vec{V}||_2, 
\end{split}
\end{equation}
where $||\cdot ||_F$ is the Frobenius norm and $|| \cdot ||_2$ is the two-norm. Therefore, let  
\begin{equation}	
\Delta_{MV} \equiv || M^{-1} ||_F ||\delta \vec{V}||_2 + ||M^{-1}||_F^2 ||\delta M||_F ||\vec{V}||_2,
\end{equation}
then 
\begin{equation}
\begin{split}	
D_I\left(\ket{v^{N}_{\rm ideal}}, \ket{v^{N}}\right)
&\lesssim  T \sqrt{||W||_{\rm max}} ||\Delta_{MV}||_{\rm max}.
\end{split}
\end{equation} 
The value $\Delta_{MV}$ depends on how to estimate $M$ and $\vec{V}$ as we see in the following. 

\subsubsection*{The infidelity in the case of using naive strategy}
In case we naively measure each term of $M$ and $V$ one by one, from the central limit theorem, 
\begin{equation}
	||\delta M||_F \simeq \frac{\sqrt{\sum_{k\ell}(\Delta M_{k\ell}^{\rm naive})^2}}{\sqrt{\numnaive}},~||\delta \vec{V}||_2 \simeq \frac{\sqrt{\sum_k (\Delta V_k^{\rm naive})^2}}{\sqrt{\numnaive}},
\end{equation}
where we use \eqref{eq:naive-variance}.
Therefore, we obtain 
\begin{equation}
	\Delta_{MV} \simeq \frac{\Deltanaive}{\sqrt{\numnaive}},
\end{equation}
where 
\begin{equation}
		\Deltanaive = || M^{-1} ||_F \sqrt{\sum_k (\Delta V_k^{\rm naive})^2} 
	+ ||M^{-1}||_F^2 ||\vec{V}||_2 \sqrt{\sum_{k\ell}(\Delta M_{k\ell}^{\rm naive})^2}.
\end{equation}
Thus, we obtain 
\begin{equation}
\label{eq:bound-naive}
	D_I\left(\ket{v^{N}_{\rm ideal}}, \ket{v^{N}}\right)
\lesssim D_I^{\rm naive},~with~ D_I^{\rm naive}=T \sqrt{||W||_{\rm max}} \Deltanaive^{\rm max} / \sqrt{\numnaive}, 
\end{equation}
where $\Delta_{\rm max}^{\rm naive} \equiv ||\Deltanaive||_{\rm max}$.

\subsubsection*{The infidelity in the case of using the shadow-based strategy}
Similarly, in case of shadow-based measurements, we obtain
\begin{equation}
		||\delta M||_F \simeq \frac{\sqrt{\sum_{k\ell}(\Delta M_{k\ell})^2}}{\sqrt{N_{\rm shot}}},~||\delta \vec{V}||_2 \simeq \frac{\sqrt{\sum_k (\Delta V_k)^2}}{\sqrt{N_{\rm shot}}},
\end{equation}
from the central limit theorem, 
where we use \eqref{eq:shadow-based-variance-m} and \eqref{eq:shadow-based-variance-v}. Thus, 
\begin{equation}
	\Delta_{MV} \simeq \frac{\Deltashadow}{\sqrt{N_{\rm shot}}},
\end{equation}
where 
\begin{equation}
		\Deltashadow = || M^{-1} ||_F \sqrt{\sum_k (\Delta V_k)^2} 
	+ ||M^{-1}||_F^2 ||\vec{V}||_2 \sqrt{\sum_{k\ell}(\Delta M_{k\ell})^2}.
\end{equation}
As a result, 
\begin{equation}
\label{eq:bound-shadow}
	D_I\left(\ket{v^{N}_{\rm ideal}}, \ket{v^{N}}\right)
\lesssim  D_I^{\rm shadow} ~with~D_I^{\rm shadow}=T \sqrt{||W||_{\rm max}} \Deltashadow^{\rm max} / \sqrt{N_{\rm shot}}, 
\end{equation}
where $\Delta_{\rm max}^{\rm shadow} \equiv \max_{0\leq t \leq T}( \Deltashadow)$.
From the equations \eqref{eq:bound-naive} and \eqref{eq:bound-shadow}, we see that $D_I^{\rm naive}$ = $D_I^{\rm shadow}$ holds when
\begin{equation}
\label{eq:num-measurement-ratio}
	\frac{N_{\rm shot}}{\numnaive} \simeq \left(\frac{\Deltashadow^{\rm max}}{\Deltanaive^{\rm max}}\right)^2,  
\end{equation}
where we assume that the value of $||W||_{\rm max}$ is almost the same in the naive strategy and the shadow-based strategy.

\subsubsection*{The infidelity in the case of using the hybrid strategy}
In case of the hybrid strategy, where we use in case $\nbb=1$, 
the number of measurements for estimating each $M_{k\ell}$ can be smaller than that for estimating $V_k$ (=$\ns$); we set the number of measurements as $\ns \alpha$ with $0 < \alpha < 1$. Then, in the hybrid strategy, 
\begin{equation}
		||\delta M||_F \simeq \frac{\sqrt{\sum_{k\ell}(\Delta M_{k\ell}^{\rm naive})^2}}{\sqrt{N_{\rm shot}\alpha}},~||\delta \vec{V}||_2 \simeq \frac{\sqrt{\sum_k (\Delta V_k)^2}}{\sqrt{N_{\rm shot}}},
\end{equation}
from the central limit theorem. We obtain
\begin{equation}
	\Delta_{MV} \simeq \frac{\Deltahybrid}{\sqrt{\ns}},
\end{equation}
where 
\begin{equation}
\Deltahybrid = || M^{-1} ||_F \sqrt{\sum_k (\Delta V_k)^2} 
	+ ||M^{-1}||_F^2 ||\vec{V}||_2 \sqrt{\frac{\sum_{k\ell}(\Delta M_{k\ell}^{\rm naive})^2}{\alpha}} .
\end{equation}
Then, 
\begin{equation}
\label{eq:bound-hybrid}
	D_I\left(\ket{v^{N}_{\rm ideal}}, \ket{v^{N}}\right)
\lesssim  D_I^{\rm hybrid} ~with~D_I^{\rm hybrid}=T \sqrt{||W||_{\rm max}} \Deltahybrid^{\rm max} / \sqrt{N_{\rm shot}}, 
\end{equation}
where $\Delta_{\rm max}^{\rm hybrid} \equiv \max_{0\leq t \leq T}( \Deltahybrid)$.
From the equations \eqref{eq:bound-naive} and \eqref{eq:bound-hybrid}, we can show that $D_I^{\rm naive}$ = $D_I^{\rm hybrid}$ holds when
\begin{equation}
\label{eq:num-measurement-ratio-hybrid}
	\frac{N_{\rm shot}}{\numnaive} \simeq \left(\frac{\Deltahybrid^{\rm max}}{\Deltanaive^{\rm max}}\right)^2,  
\end{equation}
where we assume that the value of $||W||_{\rm max}$ is almost the same in the naive strategy and the hybrid strategy.

\section{Derivation of the Haar averages}
\label{section:haar-integral}
In this section, we derive results of the Haar integration. We use the following itegration formulae \cite{Collins2006}:
\begin{align}	
\label{eq:haar-two-integral}
	\int_{\rm Haar} dU {\rm Tr}(A UBU^{\dagger}) &= \frac{{\rm Tr}(A){\rm Tr(B)}}{N}, \\
\label{eq:haar-four-integral}
		\int_{\rm Haar} dU {\rm Tr}(A UBU^{\dagger}) {\rm Tr}(C UDU^{\dagger}) &= \frac{{\rm Tr}(A){\rm Tr}(B){\rm Tr}(C){\rm Tr}(D) + {\rm Tr}(AC){\rm Tr}(BD)}{N(N+1)}, \\
\label{eq:haar-eight-integral}
	\int_{\rm Haar} dU \left({\rm Tr}(AU|0\rangle\langle 0|U^{\dagger})\right)^4 &= \frac{6{\rm Tr}(A^4) + 3[{\rm Tr}(A^2)]^2}{N^4 + 6 N^3 + 11 N^2 + 6N},
\end{align}
where $N$ is the dimension of the Hilbert space of $U$.
\subsubsection*{Derivation of \eqref{eq:vk-haar-integral}}
By using \eqref{eq:haar-two-integral} with $A=\tilde{P}_r\tilde{P}_{r^{\prime}}$ and $B=|0\rangle\langle 0|$, we obtain
\begin{equation}
\begin{split}	
\langle\vki \rangle_{\rm Haar} &= \sum_{r=1}^{\nba} \frac{(\gv)^2}{\qfunc{P_r}} +  \int_{\rm Haar} dU\sum_{r = 1}^{\nba}\sum_{r^{\prime}\neq r} \gv  \gvprime \gpp{\rm Tr}(\tilde{P}_r \tilde{P}_{r^{\prime}}  U|0\rangle\langle 0| U^{\dagger}), \\
&= \sum_{r=1}^{\nba} \frac{(\gv)^2}{\qfunc{P_r}} + 
\sum_{r = 1}^{\nba}\sum_{r^{\prime}\neq r} \gv  \gvprime \gpp \frac{{\rm Tr}(\tilde{P}_r\tilde{P}_{r^{\prime}})}{2^{n+1}}
.
\end{split}
\end{equation}
Since $\tilde{P}_r \neq \tilde{P}_{r^{\prime}}$, it holds $\tilde{P}_r \tilde{P}_{r^{\prime}} \neq I$, and therefore, ${\rm Tr}(\tilde{P}_r \tilde{P}_{r^{\prime}}) = 0$. As a result, 
\begin{equation}
\langle\vki \rangle_{\rm Haar} = \sum_{r=1}^{\nba} \frac{(\gv)^2}{\qfunc{P_r}}.	
\end{equation}
By using \eqref{eq:haar-four-integral} with $A=\tilde{P}_r\tilde{P}_{r^{\prime}}$, 
$B=\tilde{P}_u\tilde{P}_{u^{\prime}}$, and $C, D = |0\rangle\langle 0|$. 
\begin{equation}
\begin{split}	
	{\rm Var}(\vki)_{\rm Haar} &= \left(\sum_{r=1}^{\nba} \frac{(\gv)^2}{\qfunc{P_r}} + 
\sum_{r = 1}^{\nba}\sum_{r^{\prime}\neq r} \gv  \gvprime \gpp {\rm Tr}(\tilde{P}_r\tilde{P}_{r^{\prime}} U|0\rangle\langle 0|U^{\dagger})\right)^2
-  \left(\sum_{r=1}^{\nba} \frac{(\gv)^2}{\qfunc{P_r}}\right)^2
	\\
&= \sumru \gv  \gvprime \gvu \gvuprime \gpp \gppu \frac{{\rm Tr}(\tilde{P}_r\tilde{P}_{r^{\prime}}) {\rm Tr}(\tilde{P}_u\tilde{P}_{u^{\prime}})
+ {\rm Tr}(\tilde{P}_r\tilde{P}_{r^{\prime}} \tilde{P}_u \tilde{P}_{u^{\prime}})
}{2^{n+1}(2^{n+1} + 1)} \\
&= \sumru \gv  \gvprime \gvu \gvuprime \gpp \gppu \frac{{\rm Tr}(\tilde{P}_r\tilde{P}_{r^{\prime}} \tilde{P}_u \tilde{P}_{u^{\prime}})}{2^{n+1}(2^{n+1} + 1)},  
\end{split}
\end{equation}
where in the last equality, we use ${\rm Tr}(\tilde{P}_r \tilde{P}_{r^{\prime}}) = 0$.

\subsubsection*{Derivation of \eqref{eq:vknaive-haar-integral}}
By using \eqref{eq:haar-four-integral}, with $A, C = \tilde{P}_r $ and 
$B, D = |0\rangle\langle 0|$, 
\begin{equation}
\begin{split}	
		\langle \gkir\rangle_{\rm Haar} &= 
		(\gv)^2 \int_{\rm Haar} dU \left( 1 - \left({\rm Tr}(\tilde{P}_r U|0\rangle\langle 0| U^{\dagger})\right)^2 \right)
		\\
		&= (\gv)^2 \left(1 - 
		\frac{
		({\rm Tr}(\tilde{P}_r))^2 + ({\rm Tr}(\tilde{P_r}^2))
		}{2^{n+1}(2^{n+1} + 1)}
		\right) \\
		&= (\gv)^2 \left(1 - \frac{1}{2^{n+1} + 1} \right).
\end{split}
\end{equation}
By using \eqref{eq:haar-eight-integral} with $A = \tilde{P}_r$, 
\begin{equation}
\begin{split}	
	{\rm Var}( \gkir) &= 
	(\gv)^4\int_{\rm Haar} dU \left(1 - \left({\rm Tr}(\tilde{P}_r U|0\rangle\langle 0| U^{\dagger})\right)^2\right)^2 - (\langle \gkir\rangle_{\rm Haar})^2
	\\
	&= (\gv)^4 \left[\frac{6{\rm Tr}(\tilde{P}_r^4) + 3[{\rm Tr}(\tilde{P}_r^2)]^2}{2^{4(n+1)} + 6\cdot2^{3(n+1)} + 11\cdot 2^{2(n+1)} + 6 \cdot2^{n+1}} + \frac{2\cdot 2^{n+1}-1}{(2^{2(n+1)} +1)^2}\right] \\
	&= \frac{ (\gv)^4 (2\cdot 2^{2(n+1)} + 8\cdot 2^{2(n+1)})}{(2^{n+1} + 1)^2(2^{n+1}+3)}.
\end{split}
\end{equation}

\end{document}